# Leveraging rapid sintering to retain metastable zirconia in copper

Wangshu Zheng [a,b]*, Xi Chen [b], Xuyang Feng [b], Aiji Zou [a], Andrew Yun Ru Ng [c], Xiaoqing Wang [d], Zhili Dong [a], Qiang Guo [b]*, Chee Lip Gan [a]*

## Affiliation

[a] School of Materials Science and Engineering, Nanyang Technological University, 639798, Republic of Singapore

[b] School of Materials Science and Engineering, Shanghai Jiao Tong University, Shanghai, 200240, P. R. China

[c] Singapore Centre for 3D Printing, Nanyang Technological University, 639798, Republic of Singapore

[d] Department of Civil Engineering, TongJi ZheJiang College, Zhejiang, 314051, P. R. China

*Corresponding author (Email): Wangshu Zheng (cs-wangshu.zheng@ntu.edu.sg); Qiang Guo (guoq@sjtu.edu.cn); Chee Lip Gan (clgan@ntu.edu.sg)

## Abstract

Cermets combining metastable ceramics and ductile metals promise superior toughness and strength. However, retaining metastability often requires high temperature sintering that coarsens microstructures and relaxes matrix constraint. Here we introduce an ultrafast high-temperature sintering (UHS) strategy to overcome this trade-off in zirconia–copper cermets. By applying joule heating at around 100°C per second to 900°C with only a 20 second hold, we obtained cermets containing up to 50 weight percent of metastable austenite in zirconia at room temperature within a fine-grained and homogeneous microstructure. The rapid sintering kinetically favors semi-thermal austenite formation while suppressing copper grain growth and matrix relaxation, thereby stabilizing the high-temperature phase and simultaneously preserving microstructural refinement. This approach offers significant potential for copper-based composites in applications such as transformation-toughening, self-healing, and crack-detection.

**Highlights**

- Pressureless UHS retains metastable tetragonal zirconia in Cu cermets.
- A 900 °C/20 s cycle yields ~50 wt.% tetragonal phase within zirconia.
- Rapid heating decouples zirconia austenitization from Cu grain growth.
- Fine Cu grains and residual compression preserve matrix constraint.
- The strategy extends metastability engineering to transition-metal matrices.

## 1. Introduction

Cermets that combine metastable ceramics with ductile metals offer a unique combination of properties for structural and thermal applications [1-3]. In such systems, the ceramic phase can provide high hardness, thermal stability and transformation capability, while the metallic phase contributes toughness, damage tolerance and processability. Among metastable ceramics, zirconia ($ZrO_2$) is particularly attractive because it undergoes reversible martensitic phase transformations under thermal or mechanical stimuli [4-7]. The tetragonal (austenite) to monoclinic (martensite) transformation is accompanied by ~5% volume expansion and ~10% shear strain [8-10], enabling stress accommodation, crack-tip shielding and transformation-induced toughening. These features have long underpinned the design of transformation-toughened ceramics, but their implementation in ceramic–metal composites remains less explored. In our previous work [11], we incorporated such metastable zirconia into aluminum via powder metallurgy and found that a strong matrix constraint can depress the ceramic's transformation temperature as predicted by phase diagrams. The resulting metastable austenite opened up new prospects for intelligent sensors/actuators [12] and transformation-induced toughening/strengthening [13, 14].

Achieving comparable metastable phase retention in transition-metal matrices (*e.g.* Cu, Ni, Fe), however, remains challenging. To retain the high-temperature tetragonal phase at room temperature, the metal matrix must provide sufficient mechanical constraint during cooling. In Cu, conventional sintering (CS) with slow heating and prolonged holding causes substantial grain growth, particle rearrangement and stress relaxation at high temperature [15]. These processes reduce the local confinement imposed on zirconia and weaken the ability of the matrix to stabilize metastable austenite. Moreover, Cu exhibits poor wetting [16] and low chemical affinity to zirconia [17], which tend to produce weak interfaces and insufficient load transfer across the ceramic–metal boundary. Consequently, even if zirconia is thermally driven into austenitic state during sintering, the high-temperature phase cannot be effectively retained on cooling and instead reverts to martensite. This coupling between austenitization, densification, grain growth and matrix relaxation makes the fabrication of metastable zirconia–Cu cermets particularly difficult using conventional thermal routes.

Here, we employed ultrafast high-temperature sintering (UHS), a non-equilibrium method with rapid heating and a short hold [18, 19], to address this issue. The central concept is to exploit the different kinetic timescales of zirconia austenitization and Cu microstructural relaxation: the martensitic-like monoclinic-to-tetragonal transformation can be activated rapidly at elevated temperature, whereas diffusional grain growth and stress relaxation in Cu require much longer time. By using a rapid thermal cycle, UHS is expected to promote zirconia austenitization while suppressing Cu coarsening, particle segregation and loss of matrix constraint. In this work, we developed kinetic and thermodynamic models to clarify this processing window, validated the concept using zirconia–copper compacts in comparison with a CS schedule, and examined the resulting phase retention, microstructural evolution, residual stress and hardness. The results establish a processing strategy for retaining metastable zirconia in a transition-metal matrix and provide insights into the design of transformation-enabled ceramic–metal composites.

## 2. Design Rationale

*Thermodynamic rationale.* The principle of our approach is to choose a thermal profile that triggers zirconia austenitization while suppressing copper coarsening. The semi-thermal monoclinic → tetragonal transformation in zirconia is primarily chemistry-driven (**Fig. 1a**) [4, 20, 21]. It is essentially displacive (low-diffusional) and becomes thermodynamically favorable above the equilibrium temperature $T_0$. Considering 6 mol.% Ce-doped zirconia particles with mean radius $\bar{r} = 1$ μm embedded in Cu, the temperature derivative of the austenitization driving force is: $\partial \Delta G_{aus}/\partial T = -\Delta S_{aus} + 3K_{Cu}(\alpha_{Cu} - \alpha_{zirconia})\Delta V$ (see **Text S1** for derivation), where $\Delta S_{aus}$ is the entropy change of austenitization (3.1 J·mol$^{-1}$·K$^{-1}$ [11]), $K_{Cu}$ is the bulk modulus of Cu (140 GPa [22]), $\alpha_{Cu}$ and $\alpha_{aus}$ are the thermal expansion coefficients of Cu and zirconia (16.5×10$^{-6}$ K$^{-1}$ [23] and 10.6×10$^{-6}$ K$^{-1}$ [11], respectively), $\Delta V$ is the transformation volumetric change (8.1×10$^{-7}$ m$^3$·mol$^{-1}$). This gives $\partial \Delta G_{aus}/\partial T$ around -5.1 J·mol$^{-1}$·K$^{-1}$. At $T$ = 1173 K (~900°C), which exceeds the equilibrium temperature $T_0$ = 1036 K (**Table S1**), the austenitization driving force $\Delta G_{aus}$ is around -3.6 kJ·mol$^{-1}$. By contrast, Cu grain growth is driven by the reduction of grain-boundary energy $\Delta G_{gb}$ (**Fig. 1b**) and, per Gibbs-Thomson effect (**Text S2**) [24, 25], $\Delta G_{gb} = 2\gamma_{gb}/\bar{R}$, where $\bar{R}$ is the mean Cu grain radius (here 0.1 μm) and $\gamma_{gb}$ is the temperature-dependent grain-boundary energy. At $T = 1173$ K, $\gamma_{gb}$ ~0.2 J·m$^{-2}$ [26, 27], yielding $\Delta G_{gb}$ ~23.0 J·mol$^{-1}$,which is two orders of magnitude smaller than that $|\Delta G_{aus}|$. With $S_{gb}$ ~5×10$^{-4}$ J·m$^{-2}$·K$^{-1}$ being the excess grain-boundary entropy change [25], the temperature derivative $\partial \Delta G_{gb}/\partial T = -2S_{gb}/\bar{R}$ (-7×10$^{-2}$ J·mol$^{-1}$·K$^{-1}$) is also small, and grain coarsening (increasing $\bar{R}$) further weakens the driving force. Hence, to achieve sufficient austenitization while avoiding copper coarsening or melting, the optimal sintering temperature lies between $T_0$ (austenitization equilibrium temperature) and $T_m$ (copper melting point, **Fig. 1c**).

*Kinetical rationale.* At a fixed temperature within this window (**Fig. 1c**), oxygen transport in zirconia is short-range [28] (**Fig. 1a**), and austenitization proceeds rapidly. The transformed austenite fraction ($X_{aus}$) can be described by a Koistinen–Marburger-type relation [29-31] (**Text S3**): $X_{aus} = 1 - \exp[-k_{aus}(T - T_0)_+]$, with $k_{aus} = 0.013$ K$^{-1}$ [11]. Thus, at $T = 1173$ K, ~80% austenite is expected, essentially time-independent (**Table S2**). Under a practical heating ramp, a fast rate (*e.g.*, 100 K·s$^{-1}$ in UHS) reaches ~90% austenitization in ~9 s, whereas conventional sintering requires ~9250 s (~2.6 h) to attain a similar level (**Fig. 1d**). In contrast, Cu grain growth is diffusion-controlled and time ($t$)-dependent [32-35]. Defining $X_{gb}$ as the fraction of grain-boundary Gibbs free energy dissipated, a mean-field model (**Text S4**) gives: $X_{gb}(t) = 1 - (1 + K/\bar{R}^2 \cdot t)^{-1/2}$, where $K$ is an effective diffusion coefficient (1.2×10$^{-16}$ m$^2$·s$^{-1}$ [36-38]). Even at 1173 K, achieving 90% grain growth (*e.g.*, from 0.1 μm to 1 μm) requires $10^3$–$10^4$ s (several hours, consistent with conventional sintering), *i.e.*, 3-4 orders of magnitude slower than UHS-driven austenitization (**Fig. 1d**). Of note, a brief hold at peak temperature remains necessary to complete densification and preserve matrix constraint so that the metastable austenite is retained on cooling. Consequently, a high peak temperature combined with a short hold promotes rapid austenitization while minimizing Cu grain growth and stress relaxation. The resulting fine, rigid matrix mechanically stabilizes a portion of the high-temperature tetragonal phase upon cooling—this framework motivates the UHS strategy adopted here.

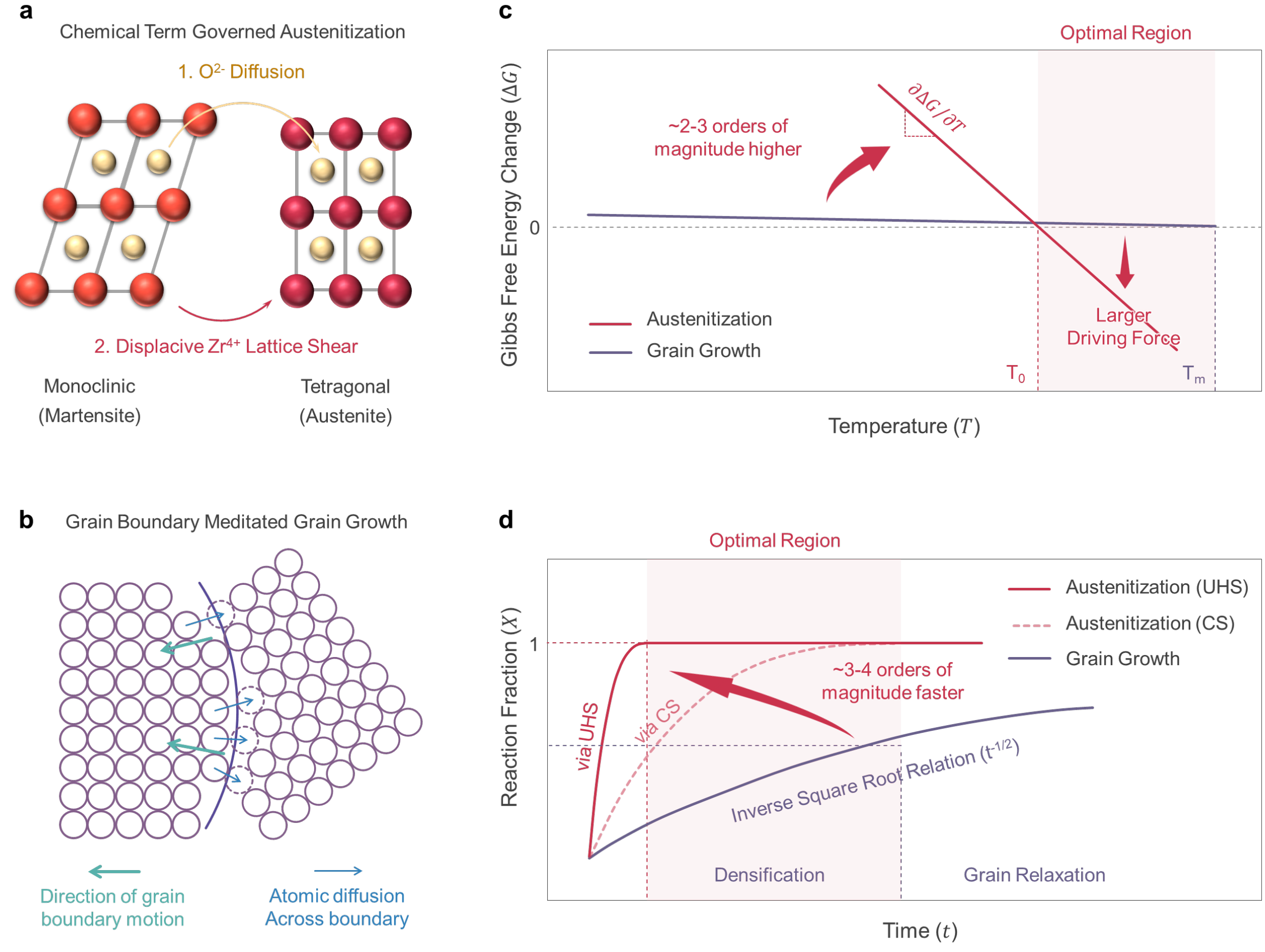

**Fig. 1 Rationale for ultrafast high-temperature sintering (UHS) in this study.** (a-b) Schematic illustration of (a) zirconia austenitization (monoclinic → tetragonal transformation) process dictated by chemistry terms, featuring the characteristics of semi-thermal martensitic transformation, and (b) copper grain growth meditated by reduction and migration of grain boundary. (c) Schematic Gibbs free energy change ($\Delta G$) of zirconia austenitization and copper grain growth as a function of temperature ($T$). $T_0$: equilibrium temperature of austenitization, $T_m$: melting point of copper. (d) Schematic reaction fraction ($X$) versus time ($t$) for austenitization of zirconia and diffusional grain growth of copper. Note: UHS – ultrafast high-temperature sintering with high ramping rate; CS – conventional sintering with low ramping rate.

## 3. Experimental

### *3.1. Materials and fabrication*

Plasma-atomized Cu powder (15-53 µm, 99.9% purity) and co-sintered $Ce_{0.06}Zr_{0.94}O_2$ powder (6CZ, ~2 µm; **fig. S1** and **Table S1**) were used as starting materials. Cu and 20 wt.% 6CZ powders were co-milled (300 rpm, 12 h) under high-purity Ar in a planetary ball mill (hardened steel media). The as-milled composite powder (**fig. S2**) was then cold-pressed at 1.0 GPa in an 8 mm die for 1 min to form green compacts. Sintering was conducted at various peak temperatures using two methods.

*Ultrafast high-temperature sintering (UHS)*: A custom DC joule-heating setup (inside a 99.99% Ar glovebox) was used (**Fig. 2a$_1$**). The details can be found in **Text S5** or our previous works [39, 40]. Each compact was sandwiched between two carbon fabric strips (**Fig. 2a$_2$**) that acted as resistive heating elements. A current ramp of ~1 A/s (~100°C/s heating rate) was applied up to the target temperature (700–1050°C), held for 20 s, then cut off (**Fig. 2a$_3$**). For select specimens, the hold time at 900°C was varied (0, 10, 20, 40, 60 s) to study its effect. The UHS-treated compacts are hereafter denoted as 6CZ–Cu (**Fig. 2a$_4$**). For comparison, cermets with undoped zirconia (0CZ, *i.e.* pure zirconia) were prepared via the same milling route (**fig. S2**) and sintered under identical UHS conditions (denoted 0CZ–Cu).

*Conventional sintering (CS)*: As a control, green compacts were sintered in a tube furnace under flowing Ar with a heating rate of 10°C/min, a hold of 1 h at the target temperature, and furnace cooling (~10°C/min). The UHS and CS thermal profiles are compared in **Fig. 2b**.

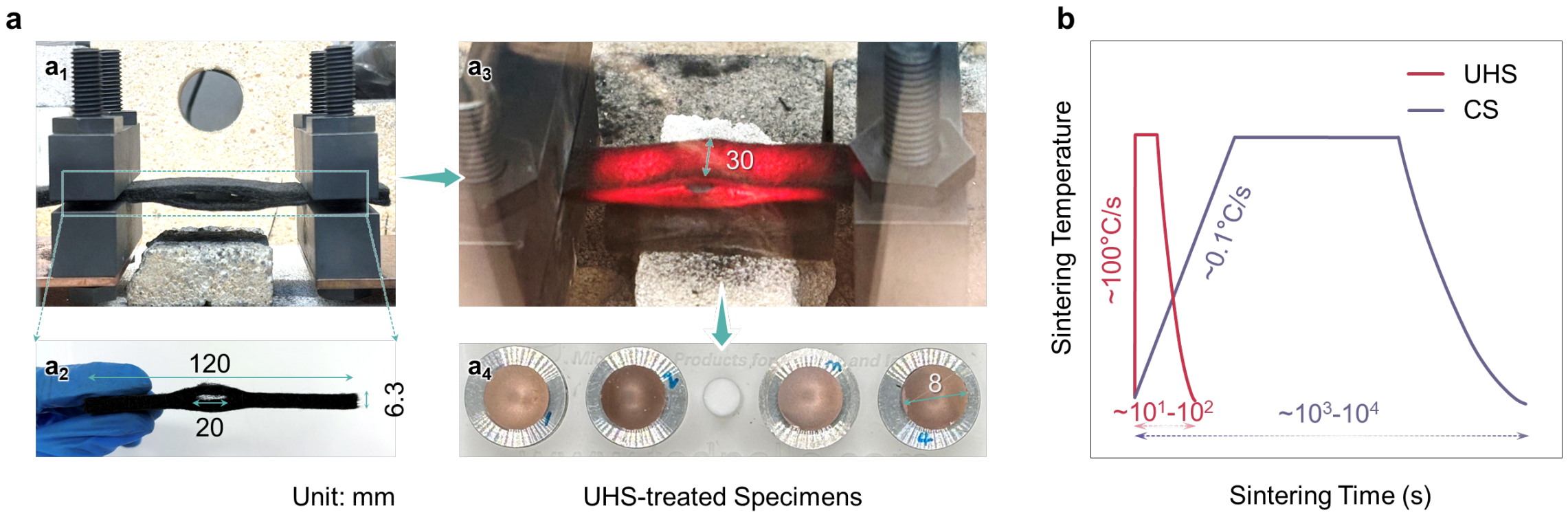

**Fig. 2 Setup, thermal profiles and austenitization behaviors of cermets from UHS and conventional sintering (CS).** ($a_1$-$a_4$) Images of ($a_1$) custom-built UHS setup, ($a_2$) carbon-fabric strip, ($a_3$) snapshot of the UHS setup during sintering, and ($a_4$) representative UHS-treated specimens. (b) Heating profiles for UHS versus CS.

### *3.2. Characterization techniques*

Microstructures were examined by scanning electron microscopy (SEM, Thermo Fisher

Helios 5, with EDS for elemental mapping). X-ray diffraction (XRD, Bruker D8 Advance, Cu Kα, λ = 1.5418 Å) was performed from 2θ = 27°–42° at 2.5°/min. The weight fraction of retained tetragonal phase was calculated using the standard intensity ratio method [41]: $f_T = I(101)_T/[I(11\bar{1})_M + I(111)_M + I(101)_T]$, where $I$ denotes the integrated intensity of the tetragonal (101) peak or the monoclinic (111) and $(11\bar{1})$ peaks. Raman spectra (WITec Alpha300R, 488 nm excitation, 1800 g/mm) were collected to assess phase-related stress. Vickers hardness (Buehler Wilson VH1150) was measured with a 5 kgf load and 10 s dwell; reported hardness values are averages of five indents per specimen.

## 4. Results and Discussion

### *4.1. Austenitization behaviors*

**Fig. 3a** shows the retained tetragonal (austenite) fraction in as-fabricated UHS-treated 6CZ–Cu (20 s hold) as a function of peak temperature. The tetragonal fraction rises from ~5% at 700°C to ~51% at 900°C, then falls to ~10% at 1050°C. Thus, ~900°C is the optimal UHS temperature for maximizing austenitization in this system. By contrast, the CS-treated 6CZ–Cu exhibits nearly complete conversion to monoclinic (virtually 0% tetragonal). This stark difference validates our model predictions (**Fig. 1**). **Fig. 3b** depicts the tetragonal fraction in 6CZ–Cu (UHS at 900°C) as a function of hold time. A 20 s hold produces the highest tetragonal fraction (~51%), whereas both shorter and longer holds result in lower fractions. Notably, even with a 60 s hold, UHS yields significantly more tetragonal phase than any of the 1 h CS conditions, underscoring the critical role of the rapid thermal cycle in preserving metastability. We also found that UHS processing of the 0CZ–Cu resulted in virtually no tetragonal phase (**fig. S4**), similar to the CS-treated 6CZ–Cu. 0CZ has a much higher equilibrium transformation temperature (~1100°C, **Table S1**), so without Ce stabilizer, the tetragonal phase cannot be retained under these sintering conditions. This confirms that the Ce dopant (6 mol%) and the rapid thermal profile act in concert to achieve partial austenitization in 6CZ–Cu.

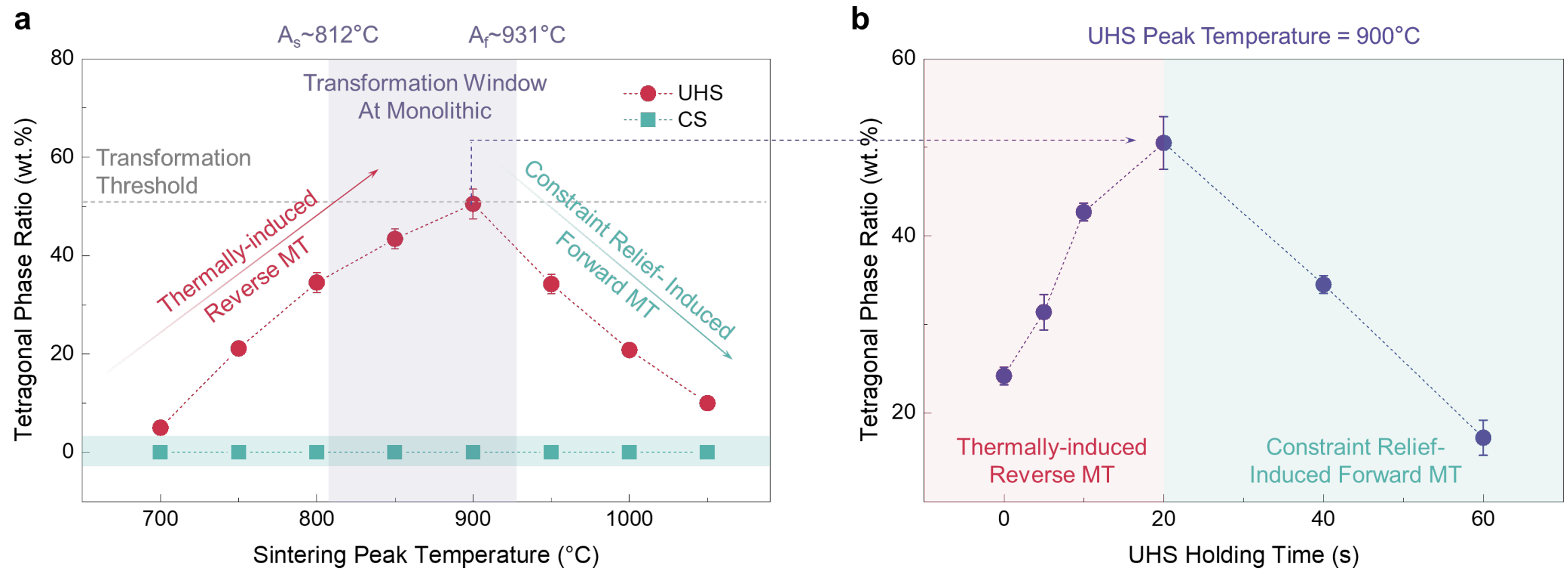

**Fig. 3 Austenitization behaviors of cermets from UHS and CS.** (a) X-ray diffraction (XRD) measured

tetragonal phase weight fraction in 6CZ–Cu as a function of sintering peak temperature for UHS (20 s hold) and CS (1 h hold). (b) Tetragonal phase fraction in UHS-treated 6CZ–Cu (at 900°C) as a function of hold time. Note: MT – martensitic transformation; $A_s$/$A_f$ – austenite start/finish temperature.

### 4.2. Microstructures

**Fig. 4a** compares backscattered-electron micrographs of 6CZ–Cu sintered by UHS at different peak temperatures (20 s hold). Clearly, residual porosity visible at 800°C peak UHS temperature is eliminated beyond 850°C, achieving full densification with clean Cu/6CZ interfaces up to 1000°C. **Fig. 4b** quantifies the Cu grain sizes, which grow from ~0.4 μm at 800°C to ~0.9 μm at 900°C, and to ~2.1 μm at 1000°C. At 900°C (UHS, **Fig. 4c**), the 6CZ particles remain uniformly dispersed throughout the Cu matrix, as corroborated by EDS mapping of Cu, Zr, and O. These traits are also ubiquitously found in specimens treated at other UHS peak temperatures (**fig. S5**). In contrast, the CS-treated 6CZ–Cu develops extensive particle clustering and segregation along Cu grain boundaries during the prolonged hold (illustrated at 900°C in **Fig. 4d, e**). In CS, the hour-long hold allows Cu grains to coarsen to ~9 μm at 900°C (**fig. S6**)—about an order of magnitude larger than the grain size in the UHS specimen.

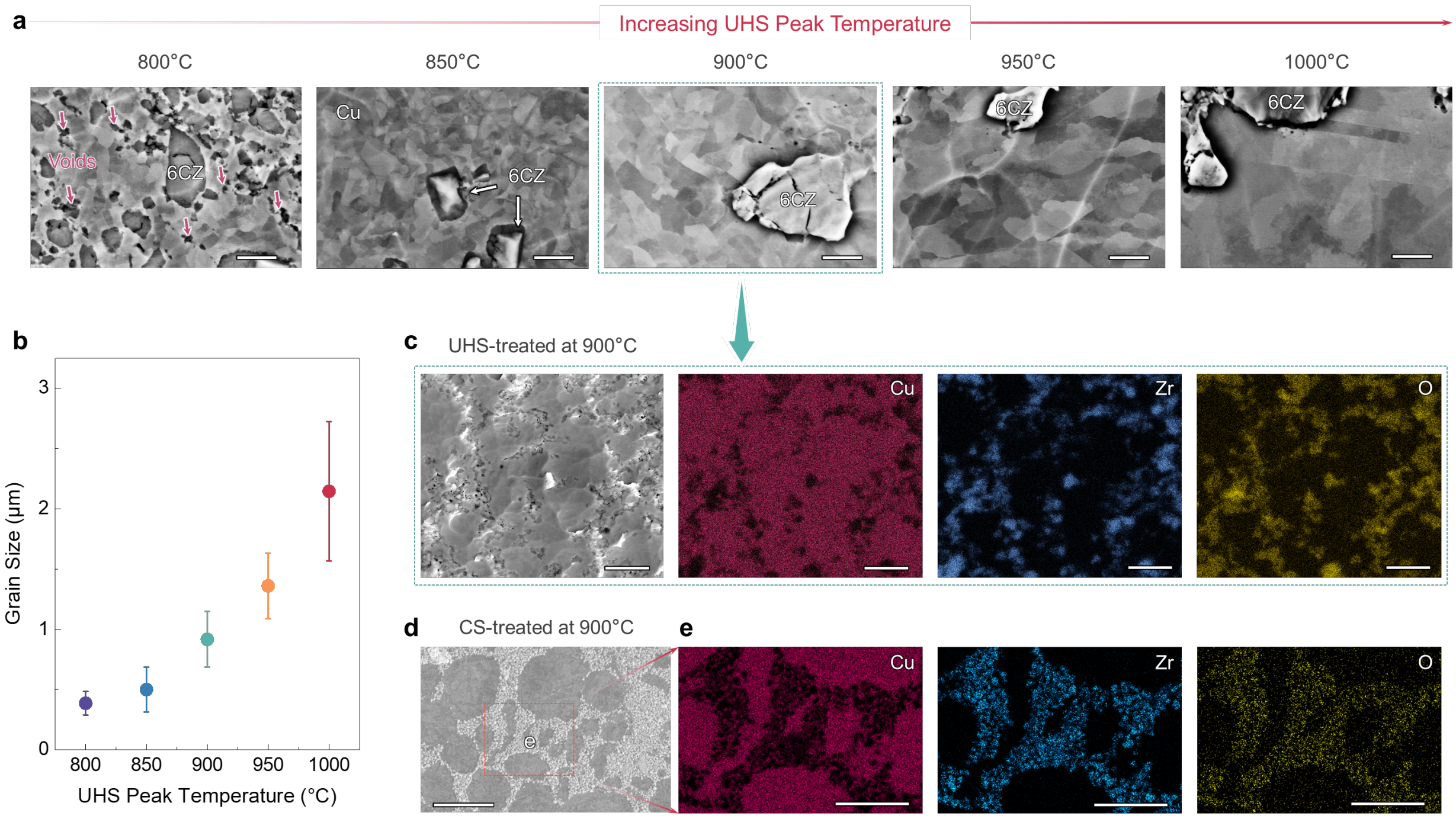

**Fig. 4 Microstructures of 6CZ–Cu cermets with varying UHS peak temperatures (20 s hold).** (a) Back-scattered electron (BSE) images showing representative Cu grains. The voids are pointed by the red arrows. (b) Statistical Cu grain size distribution in UHS-treated 6CZ–Cu with varying peak temperature. (c) Representative SEM image of UHS-treated 6CZ–Cu at 900°C, and its corresponding energy dispersive spectroscopy (EDS) color mapping of Cu (red), Zr (blue), and O (yellow). (d) Representative SEM image of CS-treated 6CZ–Cu at 900°C (1 h hold), showing the network structures. The boxed region displays (e)

the corresponding EDS color mappings. Scale bars: (a) 1 μm; (c) 5 μm; (d) 50 μm; (e) 25 μm.

### *4.3. Residual stress and hardness*

**Fig. 5a** displays Raman spectra ($E_g$ band) of tetragonal zirconia in UHS-treated 6CZ–Cu, reflecting the symmetrical transverse vibrations of oxygen atoms within the a-b plane. As the sintering temperature increases from 800°C to 900°C, the tetragonal peak shifts to higher wavenumber, indicating a build-up of average compressive stress in Zr-O bonds within 6CZ lattice; at 1000°C, the peak shifts back to lower wavenumber, a sign of stress relief. Using a piezo-spectroscopic calibration [42], the stress change ($\Delta\sigma$) is related to the peak shift ($\Delta\upsilon$) by: $\Delta\upsilon = \Pi\Delta\sigma$, where $\Pi = 2.01 \pm 0.07$ cm$^{-1}$·GPa$^{-1}$. As shown in **Fig. 5b**, the tetragonal zirconia experiences increasing compressive stress: from ~0 GPa at 800°C (reference) to ~0.59 GPa at 900°C. Beyond 900°C, the internal stress drops to ~0.10 GPa at 950°C and becomes tensile (~–0.43 GPa) at 1000°C.

These trends correlate with the mechanical response of the cermets (**Fig. 5c**). Vickers hardness of 6CZ–Cu decreases from 1.14 ± 0.04 GPa (UHS at 800°C) to 1.04 ± 0.01 GPa at 900°C, and 0.88 ± 0.01 GPa at 1000°C. Similar tendency is found in the prolonged UHS holding time (**fig. S7**). Notably, at a given sintering temperature, the UHS-treated 6CZ–Cu is at least 50% harder than the CS-treated counterpart (**fig. S8**), primarily due to its finer Cu grains. Furthermore, 6CZ–Cu shows a much smaller hardness drop (–8.8%) from 800°C to 900°C than the 0CZ–Cu (–18.9%); while a larger drop (-15.4%) from 900°C to 1000°C is found in 6CZ–Cu once the matrix constraint is relieved.

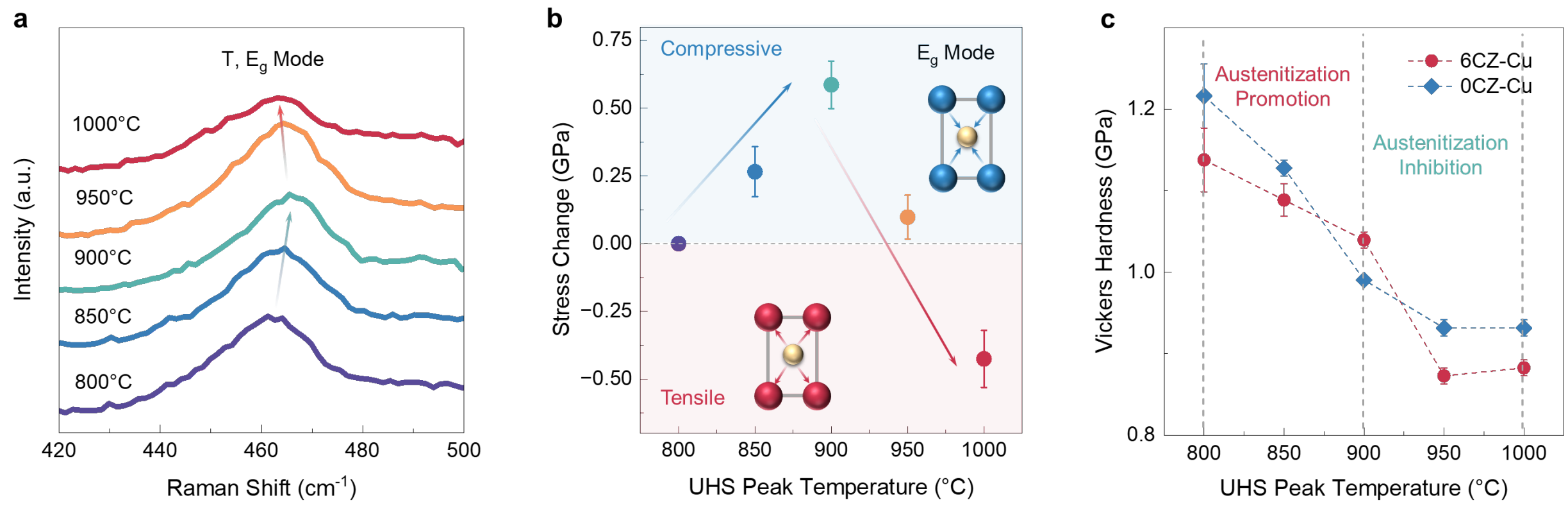

**Fig. 5 Residual stress and hardness of UHS-treated cermets.** (a) Raman spectra and (b) corresponding stress change of tetragonal phase (T, $E_g$ mode) in UHS-treated 6CZ–Cu with varying peak temperature, reflecting the symmetrical transverse vibrations (bending/lateral stretching) of oxygen atoms within the a-b plane (*i.e.* lattice distortions). (c) Hardness for 6CZ–Cu and 0CZ–Cu as a function of UHS peak temperature (20 s hold). The diagonals of indents are around 300 μm, large enough to represent the hardness of the bulks.

*4.4. Metastability mechanism*

To this end, we can uncover the combined effects of the rapid sintering schedule (**Fig. 6**). When UHS is applied above ~2/3$T_m$ of Cu (~637°C), densification (pore closure) initiates and some grain growth begins. Crucially, the rapid ramp to just above $T_0$ (~763°C for 6CZ, **Table S1**) provides a large thermodynamic driving force for austenite formation while reducing the driving force for grain growth (**Fig. 1c**). The short high-temperature exposure in UHS induces nearly instantaneous densification and then "freezes" the microstructure before significant coarsening or particle rearrangement can occur (**Fig. 4c**). Consequently, the Cu grains remain sub-micron (up to 900°C, **Fig. 4b**), in stark contrast to the multi-micron grains produced by CS (**fig. S6c$_2$**). Furthermore, a substantial compressive stress (~0.6 GPa) is retained in the tetragonal zirconia (**Fig. 5b**), sufficient to stabilize that phase upon cooling [11]. In essence, UHS minimizes matrix stress relaxation, thereby preserving a high level of constraint on the zirconia and lowering its free energy, conditions that favor retention of the metastable austenite on cooling.

If the UHS parameters are pushed too far, however, the benefits diminish. At excessively high peak temperature (*e.g.* 1000°C) or an overly long hold (*e.g.* 60 s), the process begins to resemble conventional sintering (**Fig. 6**). The matrix approaches equilibrium and internal stresses are largely released. Moreover, due to poor wetting between Cu and zirconia [16, 17], zirconia can detach and coalesce at grain boundaries when given sufficient time, leading to the pronounced clustering (**Fig. 4d**). The loss of matrix constraint, coupled with severe particle agglomeration, greatly undermines retention of metastable austenite on cooling (**Fig. 3a**). This explains the optimal UHS peak temperature and holding time to enable the highest austenitization degree, as well as the competing effect of thermally-induced austenitization and constraint relief-induced martensitic transformation.

Another intriguing observation in 6CZ–Cu is the broad austenitization temperature window (**Fig. 3a**), compared to the sharp transformation in monolithic 6CZ ceramics (**fig. S1**). In freestanding 6CZ, the martensite transforms to austenite over a relatively narrow interval between its start ($A_s$ ~812°C) and finish ($A_f$ ~931°C) temperatures. In the cermet, however, substantial austenitization is observed from ~700°C to 1050°C (**Fig. 3a**). This broad transformation window arises for two reasons. First, increasing sintering temperature has dual, opposing effects on the phase balance (**Fig. 3a**): it thermally drives austenitization while promoting matrix stress relief. These two competing effects can produce a wide transformation window rather than a sharp transition. Second, individual 6CZ experiences different local conditions. Each particle's orientation and surrounding constraint determine its effective transformation temperature [11, 43, 44]. Particles under higher local compressive stress will stay austenite at lower temperatures, whereas those in more relaxed regions transform at higher temperatures. This heterogeneity produces a spread of transformation temperatures, analogous to the continuous transformation mode [7, 12, 45]. As a consequence, even with UHS, the maximum retained austenite fraction plateaus around ~50 wt.% (**Fig. 3a, b**), unlike full austenitization achieved in aluminum in our previous works [11-13]. This threshold behavior can be understood by the heterogeneity of 6CZ, as mentioned above: the nuances of their constraint stresses and orientations make it difficult to metastabilize all the particles at a designated temperature. We also noted that adding external pressure during UHS could potentially raise this limit by providing extra confinement [46].

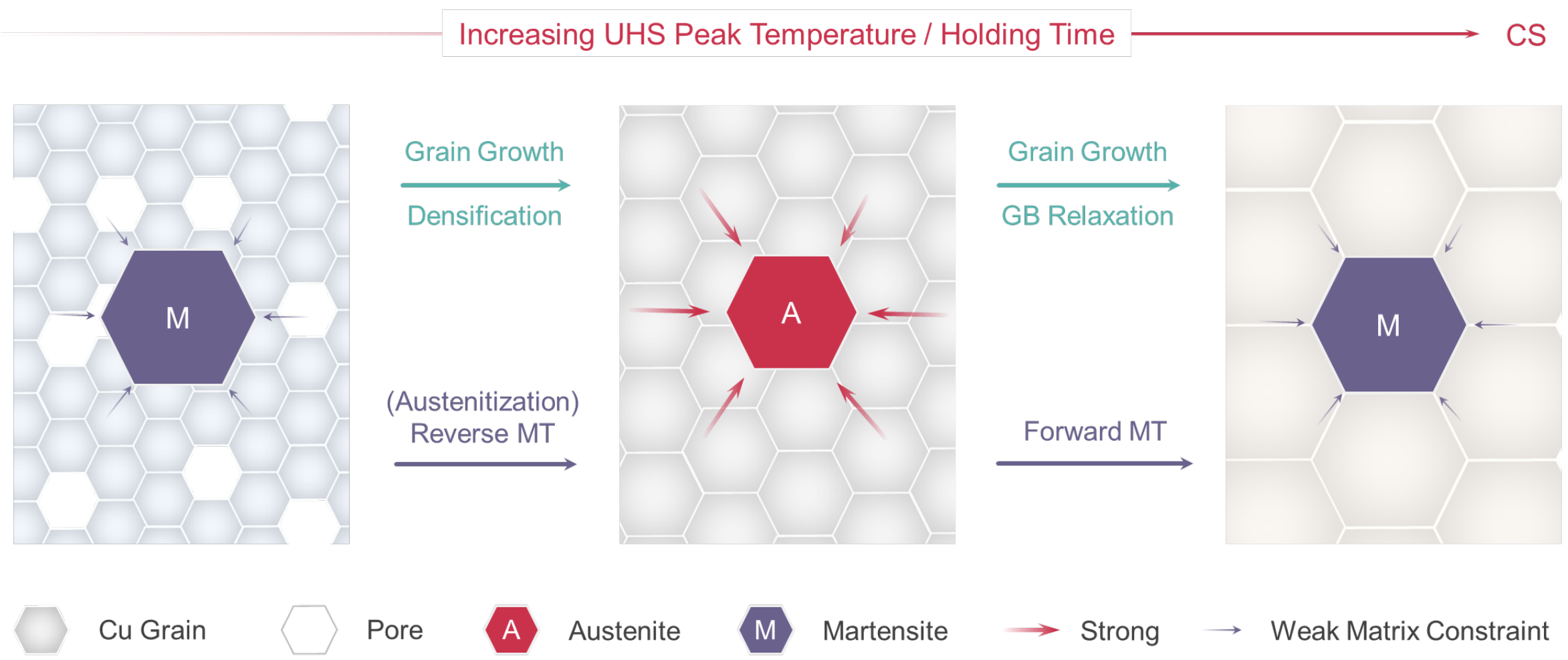

**Fig. 6 Metastability mechanism of UHS-treated cermets.** Schematic illustration of microstructural and phase evolution of zirconia-copper cermets via UHS with varying thermal profiles. CS schedule is treated as the case with elevated temperature or prolonged holding time. Note: GB – grain boundary, MT – martensitic transformation.

### *4.5. Implications*

The successful implementation of metastable zirconia–Cu cermets through a pressureless UHS route has several implications. From a processing perspective, the rapid thermal cycle provides an energy-efficient and cost-effective pathway for fabricating ceramic–metal composites while avoiding excessive grain growth, particle segregation and stress relaxation. This is particularly important for metal matrices such as Cu, where conventional sintering readily weakens the matrix constraint required to retain metastable ceramic phases.

From a mechanical perspective, the retained metastable austenite provides an additional transformation-mediated deformation mechanism. During austenitization, the ~5% volume contraction of zirconia can introduce residual compressive stresses around the ceramic particles [47], which may help stabilize the Cu matrix by impeding dislocation motion and grain-boundary migration. Under high applied stress, such as indentation, the retained tetragonal phase can further transform to martensite [48], providing local transformation toughening and contributing to the enhanced hardness observed in the UHS-treated cermets (**Fig. 5c** and **fig. S4**). Although the present study focuses primarily on hardness, this transformation capability is expected to benefit broader mechanical properties, including plasticity, tensile strength, fracture toughness and impact resistance, which merit systematic investigation in future work.

Beyond mechanical reinforcement, metastable zirconia–Cu cermets may also offer functional opportunities. The reversible tetragonal–monoclinic transformation can generate internal stress, strain accommodation and local structural changes, suggesting potential applications in self-toughening, crack mitigation, damage sensing and thermomechanical cycling environments. In addition, zirconia–copper systems are relevant to catalytic applications, where zirconia phase stability, residual stress and ceramic–metal interfacial structure may influence surface reactivity. While catalytic performance is beyond the scope of the present work, the ability to tune metastable zirconia within a Cu matrix provides a possible route for future studies on mechanically and chemically active ceramic–metal catalysts.

## 5. Conclusions

In summary, we developed a pressureless ultrafast high-temperature sintering strategy to retain metastable zirconia in a Cu matrix. Guided by kinetic and thermodynamic considerations, the rapid thermal cycle was designed to activate monoclinic-to-tetragonal austenitization in Ce-doped zirconia while suppressing Cu grain growth, particle segregation and matrix stress relaxation. By applying Joule heating at ~100 °C $s^{-1}$ to 900 °C with a 20 s hold, the tetragonal fraction within the zirconia phase reached ~50 wt.% at room temperature, whereas conventionally sintered counterparts showed nearly complete reversion to the monoclinic phase. The retained metastability is attributed to the strong matrix constraint preserved by UHS, as evidenced by the fine and homogeneous Cu microstructure, improved particle dispersion and retained compressive stress in tetragonal zirconia.

These results demonstrate that rapid thermal processing can decouple fast ceramic phase transformation from slower metallic microstructural relaxation, providing a practical route for designing metastable ceramic–metal composites. Excessively high sintering temperature or prolonged holding promoted Cu coarsening, stress relaxation and particle clustering, thereby reducing the retained austenite fraction. The ability to retain transformation-capable zirconia in Cu expands matrix-constraint engineering beyond Al-based systems and suggests future opportunities in transformation toughening, impact resistance, damage sensing and thermomechanically active cermets.

## Acknowledgment

This work was supported by the National Natural Science Foundation of China (Grant Nos. 523B2005 and 52192595). The authors thank Dr. Zehui Du and Ms. Chia Wen Soon for their valuable discussion and assistance on the setups. W.Z. appreciates the Facility for Analysis, Characterisation, Testing and Simulation (FACTS), NTU, and the National Integrated Centre for Evaluation (NiCE), NTU, for access to their characterization facilities. W.Z. also thanks

Outstanding Doctoral Graduate Development Scholarship of Shanghai Jiao Tong University.

## Competing interests

Authors declare that they have no competing interests.

## Data and materials availability

All data needed to evaluate the conclusions in this paper are provided in the main text and supplementary materials. Additional data or analysis files can be made available from the corresponding authors upon reasonable request.

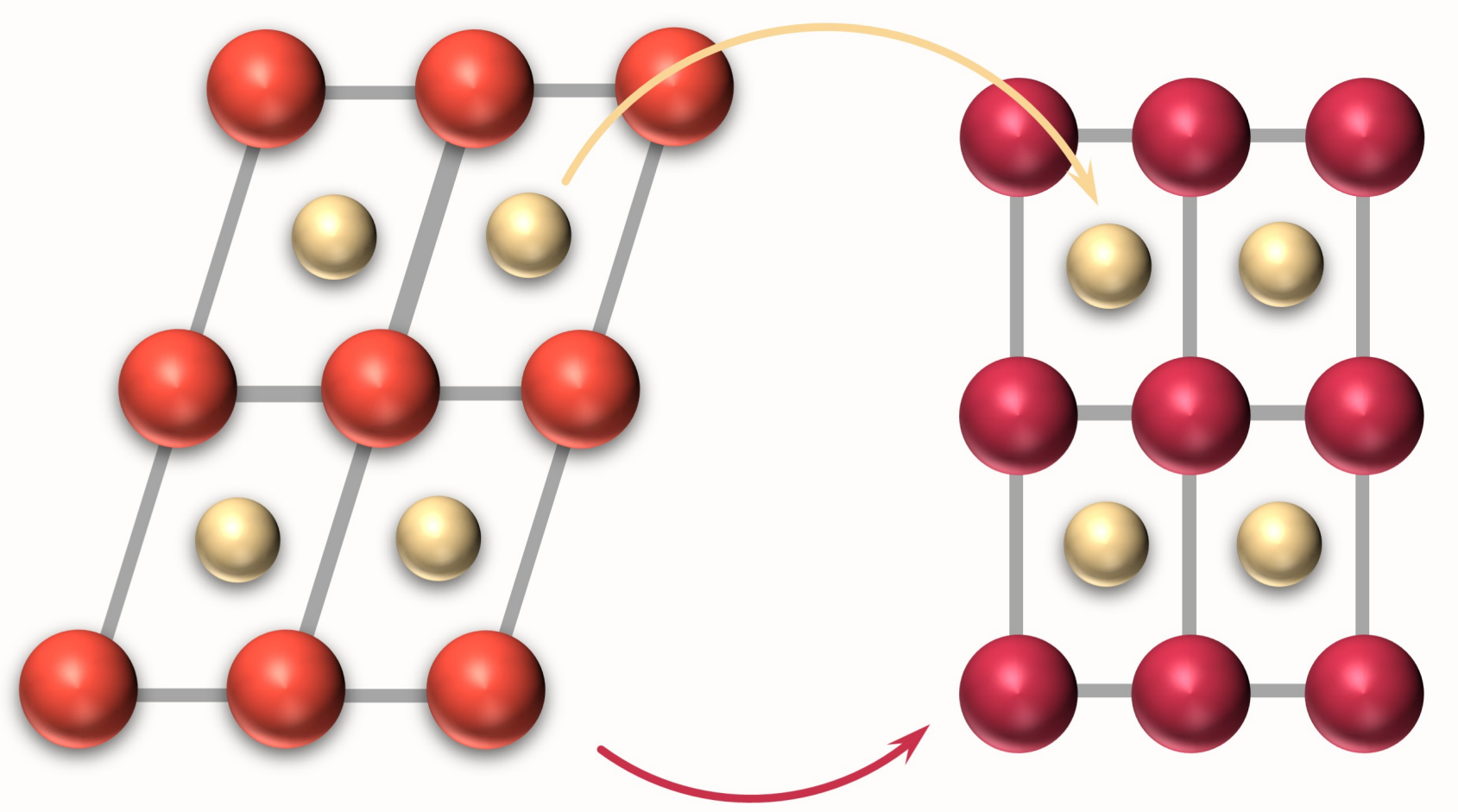
Chemical Term Governed Austenitization
1. $O^{2-}$ Diffusion
2. Displacive $Zr^{4+}$ Lattice Shear
Monoclinic
(Martensite)
Tetragonal
(Austenite)

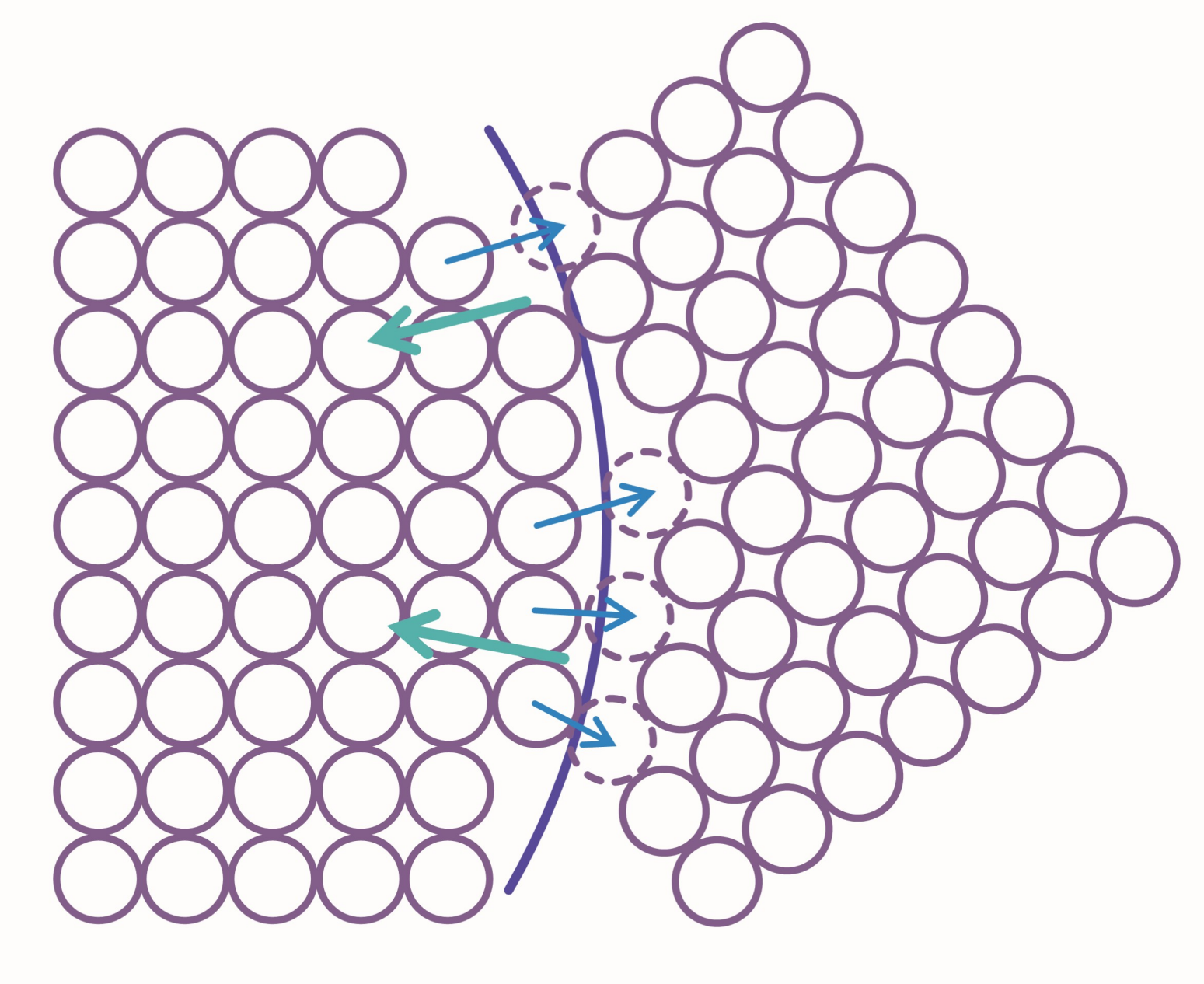
Grain Boundary Meditated Grain Growth
Direction of grain boundary motion
Atomic diffusion Across boundary

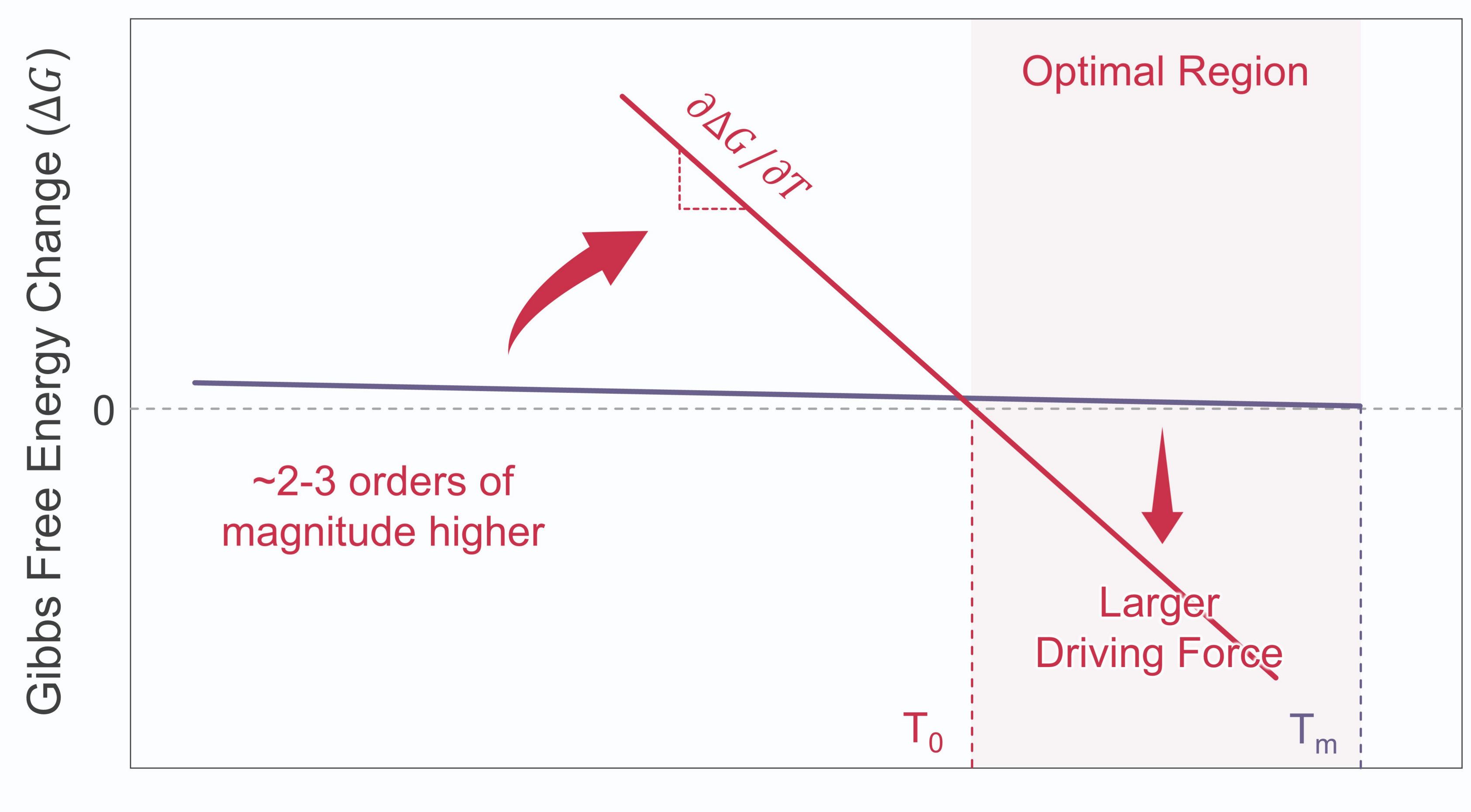
Gibbs Free Energy Change (ΔG)
Optimal Region
∂ΔG/∂T
0
~2-3 orders of magnitude higher
Larger Driving Force
T0
Tm
Temperature (T)

Austenitization (UHS)
Austenitization (CS)
Grain Growth

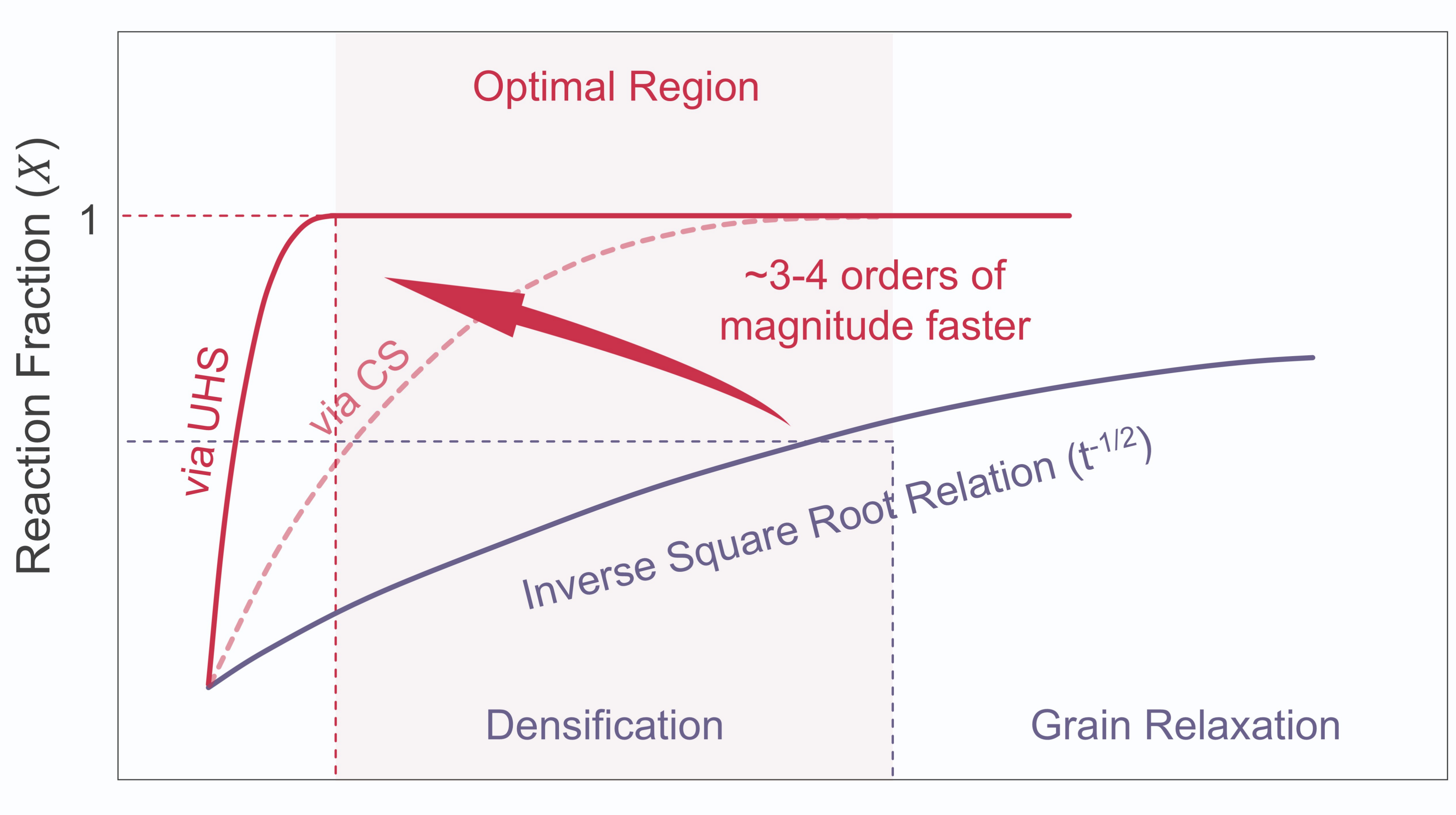
Reaction Fraction (X)
Optimal Region
1
~3-4 orders of magnitude faster
via UHS
via CS
Inverse Square Root Relation (t-1/2)
Densification
Grain Relaxation
Time (t)

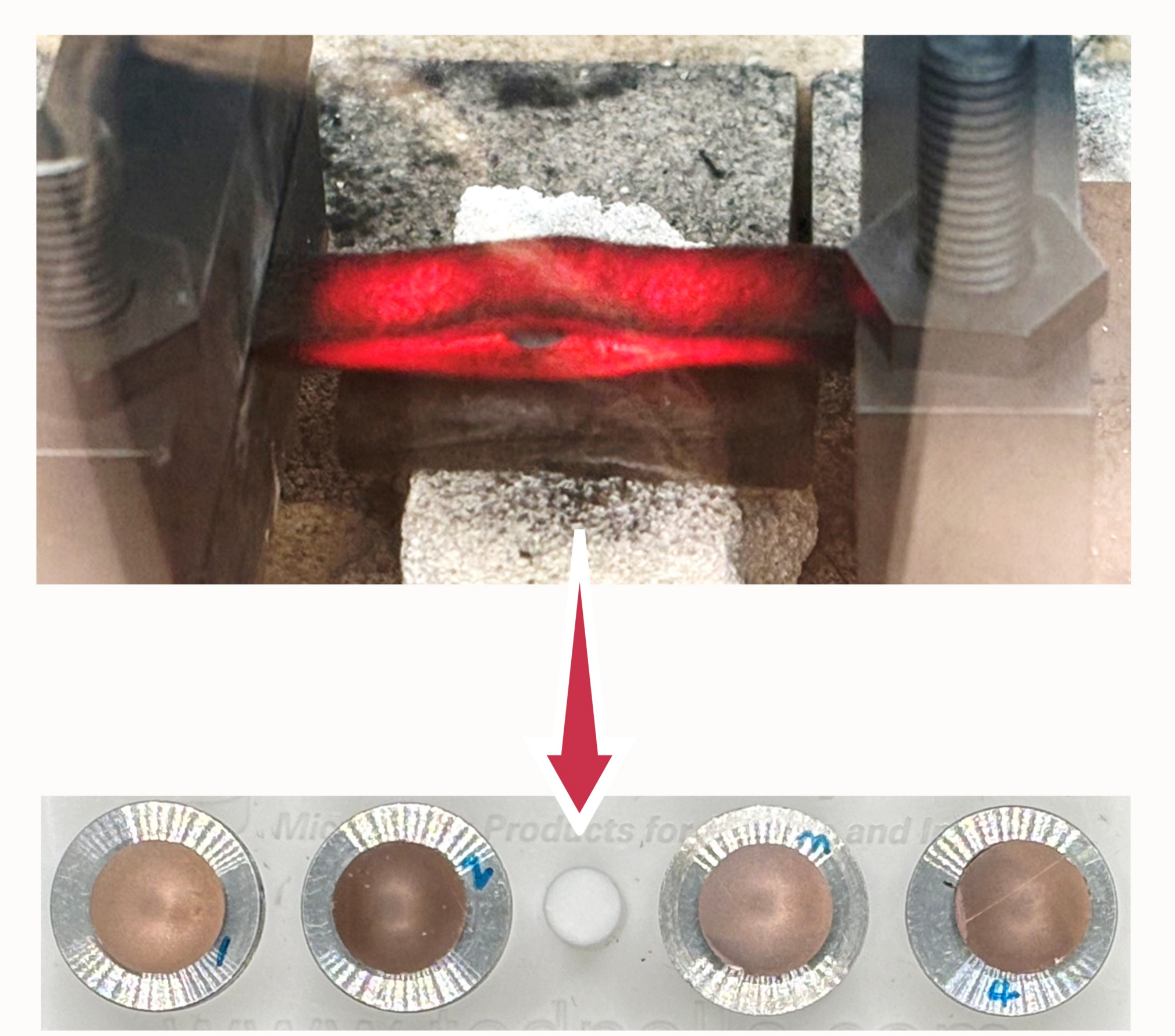

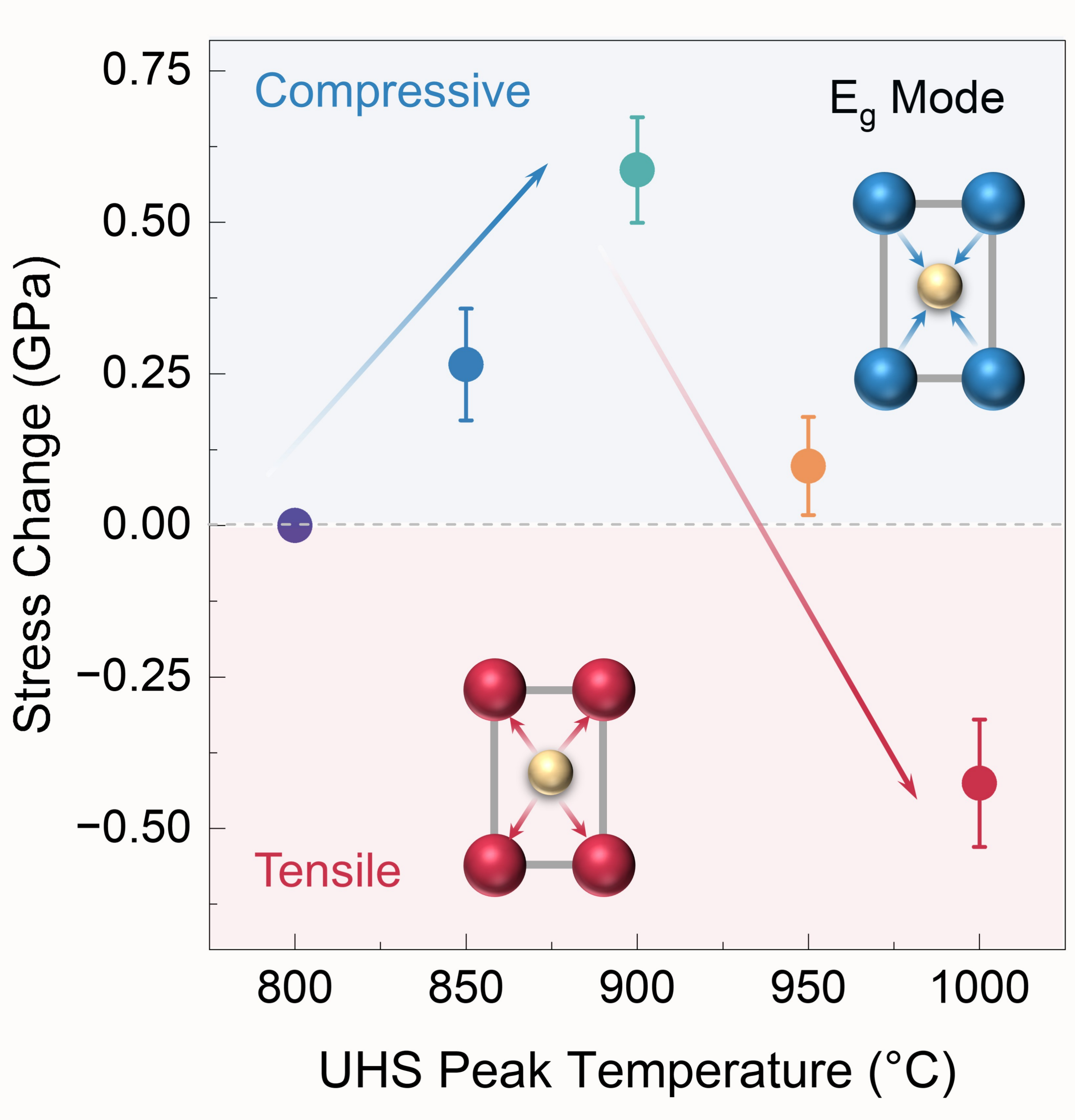
Compressive
Eg Mode
Tensile
Stress Change (GPa)
0.75
0.50
0.25
0.00
−0.25
−0.50
800
850
900
950
1000
UHS Peak Temperature (°C)

# Supplementary Information for

## Leveraging rapid sintering to retain metastable zirconia in copper

W. Zheng *et al.*

Corresponding author (Email): Wangshu Zheng (cs-wangshu.zheng@ntu.edu.sg); Qiang Guo (guoq@sjtu.edu.cn); Chee Lip Gan (clgan@ntu.edu.sg)

**This file includes**:

Supplementary Text

Figs. S1 to S7

Tables S1 to S2

References

## Supplementary Text

Text S1. Deduction of Gibbs Free Energy Change for Austenitization in Zirconia

We consider the austenitization (monoclinic → tetragonal transformation) in zirconia particles (with 6 mol.% cerium dopant) embedded in a copper (Cu) matrix. Following standard semi-thermal transformation thermodynamics [1-3], the molar Gibbs free-energy change $\Delta G_{aus}(T)$ is decomposed as a sum of chemical ($\Delta G_{chemical}$), elastic strain ($\Delta G_{elastic}$), friction ($\Delta G_{friction}$) and external mechanical terms [4, 5]:

$$\Delta G_{aus}(T) = \Delta G_{chemical} + \Delta G_{elastic} + \Delta G_{friction} + \Delta G_{external} \tag{S1}$$

For single crystalline zirconia ceramics with micrometer size, the friction energy $\Delta G_{\mathrm{Friction}}$ caused by defects and free surfaces is typically negligible [6]. Since we only consider thermally-induced austenitization, $\Delta G_{External}$ is zero due to the absence of applied stress. The chemical energy can be expressed as [4, 7]:

$$\Delta G_{chemical} = \Delta H - T\Delta S \tag{S2}$$

where $\Delta H$ and $\Delta S$ are the enthalpy change and entropy change during austenitization (3.3 $kJ \cdot mol^{-1}$, 3.1 $J \cdot mol^{-1} \cdot K^{-1}$, respectively) [8].

Assuming the volume expansion upon transformation is isotropic in three dimensions, thermal-expansion mismatch between Cu and zirconia induces a nearly hydrostatic constraint within the particle. Using a mean-field estimate,

$$\Delta G_{elastic} = P(T) \cdot \Delta V \tag{S3}$$

wherein:

$$P(T) = 3K_{Cu}(\alpha_{Cu} - \alpha_{aus})(T - T_{ref}) \tag{S4}$$

where $K_{Cu}$ is the bulk modulus of Cu (140 GPa [9]), $\alpha_{Cu}$ and $\alpha_{aus}$ is thermal expansion coefficients of Cu and zirconia ($16.5 \times 10^{-6}$ $K^{-1}$ [10] and $10.6 \times 10^{-6}$ $K^{-1}$ [8], respectively), $T_{ref}$ is the reference temperature (298 K). The transformation volumetric change of zirconia is taken as $\Delta V = \varepsilon_v V_{aus}$, with volumetric strain $\varepsilon_v$ being -0.04 [3], and molar volume $V_{aus}$ being $2.02 \times 10^{-5}$ $m^3 \cdot mol^{-1}$. Therefore, $P(T) \cdot \Delta V < 0$; *i.e.*, constraint assists austenitization upon heating. Differentiating **Eq. S1** yields:

$$\frac{d\Delta G_{aus}}{dT} = -\Delta S + 3K_{Cu}(\alpha_{Cu} - \alpha_{zirconia})\Delta V \tag{S5}$$

Therefore, $d\Delta G_{\mathrm{aus}}/dT$ is around -5.1 $J \cdot mol^{-1} \cdot K^{-1}$, indicating that $\Delta G_{aus}$ decreases monotonically with $T$. At the given temperature $T$ (*e.g.* 1173 K) above equilibrium temperature $T_0$ = 1036 K (**Table S1**), the driving force of austenitization $\Delta G_{aus}$ is calculated as around -3.6 $kJ \cdot mol^{-1}$.

Text S2. Deduction of Gibbs Free Energy Change for Grain Growth in Copper

We consider a polycrystalline Cu system with the average equiaxed grain radius $\bar{R}$ of 0.1 μm. At fixed composition and pressure, the Gibbs free energy can be decomposed into a bulk term and a grain-boundary term ($G_{gb}$). During the grain growth, the bulk term is essentially independent of the average grain radius $R$; therefore, at the sintering temperature $T$, we just focus on the grain-boundary term. According to the Gibbs-Thomson effect, polycrystalline materials have a free energy higher than that of single-crystal materials; the growth of grains is driven by this additional reduction of grain-boundary energy (or curvature of grain boundary), resulting in interfacial migration [11]:

$$\Delta G_{gb}(T,\bar{R}) = \frac{2V_{Cu}}{R} \cdot \gamma_{gb}(T) \tag{S6}$$

where $V_{Cu}$ is the molar volume of Cu ($7.09\times10^{-6}$ $m^3\cdot mol^{-1}$). $\gamma_{gb}$ is the grain-boundary energy, defined as [12]:

$$\gamma_{gb}(T) = U_{gb} - TS_{gb} \tag{S7}$$

where $U_{gb}$ is the internal energy excess per unit area, $S_{gb}$ is the entropy excess per unit area. Because grain boundaries possess significant configurational, vibrational, and magnetic disorder compared to the crystal interior, $S_{gb} > 0$ (~$5\times10^{-4}$ $J\cdot m^{-2}\cdot K^{-1}$ for Cu [12]). Therefore:

$$\frac{d\gamma_{gb}}{dT} < 0 \;\Rightarrow\; \gamma_{gb}(T) \text{ decreases with temperature} \tag{S8}$$

Over a wide temperature range below the melting point ($T_m$ = 1358 K for Cu), $\gamma_{gb}$ can be further expanded as the following equation, which decreases linearly with $T$ [13-15]:

$$\gamma_{gb}(T) = \gamma_{ref} - S_{gb}(T - T_{ref}) \tag{S9}$$

where $\gamma_{ref}$ is the grain-boundary energy (~0.6 $J\cdot m^{-2}$, [16, 17]) at a reference temperature $T_{ref}$ (298 K). Substituting **Eq. S9** into **Eq. S6** gives a compact expression for the grain-boundary Gibbs free energy:

$$\Delta G_{gb}(T,\bar{R}) = \frac{2V_{Cu}}{R}\left[\gamma_{ref} - S_{gb}(T - T_{ref})\right] \tag{S10}$$

At fixed $R$, the derivative of grain boundary free energy over temperature is therefore:

$$\left(\frac{\partial \Delta G_{gb}}{\partial T}\right)_{\bar{R}} = -\frac{2V_{Cu}}{\bar{R}} S_{gb} \tag{S11}$$

Then, we have: $\left(\partial \Delta G_{gb}/\partial T\right)_{R=0.1\ \mu m}$ equals to $-7.1\times10^{-2}$ $J\cdot mol^{-1}\cdot K^{-1}$, *i.e.*, $\Delta G_{gb}$ (driving force) decreases linearly with temperature (similar to the $\gamma_{gb}$). Representative magnitudes of the grain-boundary Gibbs free energy change $\Delta G_{gb}$ from **Eq. S5** are:

$$\begin{cases} T = 298\ \text{K:}\ \ \Delta G_{gb} \approx 85.1\ \text{J} \cdot \text{mol}^{-1} \\ T = 1173\ \text{K:}\ \ \Delta G_{gb} \approx 23.0\ \text{J} \cdot \text{mol}^{-1} \\ T = 1300\ \text{K:}\ \ \Delta G_{gb} \approx 14.0\ \text{J} \cdot \text{mol}^{-1} \end{cases} \tag{S12}$$

Therefore, at the same temperature $T$ (1173 K) above $T_0$, $d\Delta G_{\text{aus}}/dT$ is around -5.1 J·mol$^{-1}$·K$^{-1}$ (**Text S1**), a steeper slope compared to that of Cu grain growth (-7.1×10$^{-2}$ J·mol$^{-1}$·K$^{-1}$ at the grain radius of 0.1 μm). Additionally, the driving force of austenitization $\Delta G_{aus}$ is around -3.6 kJ·mol$^{-1}$, 2 orders of magnitude higher than $\Delta G_{gb}$ (23.0 J·mol$^{-1}$).

**Eq. S11** also makes explicit the $1/\bar{R}$ dependence of grain-boundary Gibbs free energy change. The thermodynamic driving force, expressed directly as a potential (per-mole) gradient with respect to grain size at fixed $T$ is:

$$\left(\frac{\partial \Delta G_{gb}}{\partial \bar{R}}\right)_T = -\frac{2V_{Cu}}{\bar{R}^2}\gamma_{gb}(T) < 0 \tag{S13}$$

which states that increasing $\bar{R}$ reduces the grain-boundary Gibbs free energy change (driving force). For instance, with the grain coarsening of $\bar{R}$ to 1 μm, $\left(\partial \Delta G_{gb}/\partial T\right)_{R=1\ \mu m}$ equals to -7.1×10$^{-3}$ J·mol$^{-1}$·K$^{-1}$, an even smaller driving force for Cu grain growth.

It should be noted that **Eqs. S3–S8** provide a purely thermodynamic description of the driving force for Cu grain growth expressed directly as Gibbs free energy (per mole). The formulation assumes (i) isotropic grain-boundary energy, (ii) equiaxed topology with a single radius parameter $R$, (iii) negligible stored-energy differences between grains, and (iv) absence of solute drag or second-phase pinning. Kinetic effects (*e.g.*, boundary mobility) are intentionally excluded and will be discussed later; actual growth rates follow from kinetics multiplying the thermodynamic potential, but they do not alter the free-energy deductions above.

## Text S3. Deduction of Reaction-Fraction Kinetics for Austenitization in Zirconia

In the reverse martensitic transformation (austenitization), migration of the $Zr^{4+}$ and $O^{2-}$ ions to their respective lattice sites is short-range diffusion-controlled [18], disfavoring the burst-like behavior. Within each parent martensite phase, the new austenite nucleus can be supersaturated; once a relatively small activation energy is reached, the nucleation and growth can be triggered and finished in a short time [19, 20]; therefore Johnson–Mehl–Avrami–Kolmogorov (JMAK) model is not applicable in this process. Instead, as raised in our previous work [8], semi-empirical Koistinen-Marburger (K-M) equation [21-23] is used to derive the austenite fraction transformed $X_{aus}$:

$$X_{aus} = 1 - \exp[-k_{aus}(T - T_0)_+] \tag{S14}$$

where $k_{aus}$ is the semi-thermal "sensitivity" parameter (0.013 $K^{-1}$ [8]). For simplicity, the matrix (Cu) constraint induced entropy change is incorporated therein. As listed in **Table S2**, the fraction transformed jumps as soon as the temperature crosses a threshold (equilibrium temperature $T_0$ ~1036 K) and then follows a simple exponential in temperature. For instance, the sintering temperature 1173 K gives rise to 83% transformed austenite. Of particular note, the transformed fraction under different matrix constraint may vary. Here, $X_{aus}$ is estimated based on the matrix constraint stress of -827 MPa obtained from our previous work in aluminum matrix [8].

On the other hand, it is apparent that time ($t$) does not control the fraction $X_{aus}$, yet thermal driving force (temperature $T$) does. In practice, temperature rises with a certain ramping rate $\dot{T}$. For instance, at a relatively fast ramping rate (*e.g.* 100 K/s), which features ultrafast high-temperature sintering (UHS), it takes only 9 s to reach 91% austenitization; by contrast, under conventional sintering, it takes 9250 s (~2.6 h) at similar austenitization degree.

Text S4. Deduction of Reaction-Fraction Kinetics for Grain Growth in Copper

As noted in **Text S2**, in this particular polycrystalline Cu system with the average equiaxed grain radius $\bar{R}$ of 0.1 μm, there is no solute drag effect. Meanwhile, the relatively large size of zirconia particles ($\bar{r}$ = 1 μm) allows us to neglect the Zenner pinning effect. Herein, the isothermal grain growth in Cu is mainly governed by the grain-boundary curvatures. In a mean-field description, the velocity of the grain boundary is governed by the reduction of grain-boundary energy [14, 15, 24, 25]:

$$\frac{d\bar{R}}{dt} = M_{gb}^{poly} \cdot \Delta G_{gb} \tag{S15}$$

where $M_{gb}^{poly}$ is the nominal grain-boundary mobility. According to **Eq. S6**, we have:

$$\frac{d\bar{R}}{dt} = M_{gb}^{poly} \cdot \frac{2\gamma_{gb} V_{Cu}}{\bar{R}} \tag{S16}$$

Based on the relationship raised by M. Hillert [26]:

$$M_{gb}^{poly} \approx \frac{1}{4} M_{gb} \tag{S17}$$

$$M_{gb} \approx \frac{D_{gb}}{\delta RT} \tag{S18}$$

where $D_{gb}$ is the grain boundary diffusivity, $\delta$ is the atomic jump distance in the grain boundary (taken to be $3a$, where $a$ = 0.362 nm is the lattice constant for Cu), $R$ is the gas constant (8.31 $J \cdot mol^{-1} \cdot K^{-1}$). Integrate **Eq. S16**, and substitute the **Eq. S17** and **Eq. S18** therein; then we have:

$$\bar{R}^2(t) - \bar{R}_0^2 = \left(\frac{\gamma_{gb} V_{Cu}}{\delta RT} \cdot D_{gb}\right) t \tag{S19}$$

The grain boundary diffusivity has an Arrhenius temperature dependence with an activation energy for grain boundary diffusion, $Q_{gb}$, such that:

$$D_{gb} = D_{gb}^{0}\, exp\left(-\frac{Q_{gb}}{RT}\right) \tag{S20}$$

where $D_{gb}^{0}$ = 1.84×10$^{-12}$ m$^{2}$·s$^{-1}$ and $Q_{gb}$ = 7.25 ×10$^{4}$ J·mol$^{-1}$ [27-29], were used to estimate $D_{gb}$. To simplify the equation, we define a temperature-dependent nominal diffusion coefficient $K$ as:

$$K(T) = \frac{\gamma_{gb}(T)V_{Cu}}{\delta RT} \cdot D_{gb}^{0}\, exp\left(-\frac{Q_{gb}}{RT}\right) \tag{S21}$$

Then **Eq. S22** is rewritten as:

$$\bar{R}^{2}(t) - \bar{R}_{0}^{2} = K(T)t \tag{S22}$$

following with the parabolic law. A natural, thermodynamically consistent reaction fraction is the fraction of grain-boundary Gibbs free energy removed relative to the initial state. According to **Eq. S6**, under isothermal conditions, $\gamma_{gb}(T)$ is the constant, and $\bar{R}(t)$ is obtained from **Eq. S19**. Define the "reaction fraction" as the fraction of grain-boundary Gibbs free energy released, and then we have:

$$X_{gb}(t) = 1 - \frac{\Delta G_{gb}\left(T, \bar{R}(t)\right)}{\Delta G_{gb}(T, \bar{R}_{0})} = 1 - \frac{\bar{R}_{0}}{\bar{R}(t)} = 1 - \frac{\bar{R}_{0}}{\sqrt{\bar{R}_{0}^{2} + K(T)t}} \tag{S23}$$

For instance, at the sintering temperature $T$ of 1173 K, $\gamma_{gb}(1173\ \mathrm{K})$ is 0.16 J·m$^{-2}$, and $K(1173\ \mathrm{K})$ is calculated as 1.19×10$^{-16}$ m$^{2}$·s$^{-1}$. Finally, we have:

$$\begin{cases} R = 0.2\ \mu\mathrm{m}:\ t \approx 253\ \mathrm{s} \\ R = 0.5\ \mu\mathrm{m}:\ t \approx 2021\ \mathrm{s} \\ R = 1\ \mu\mathrm{m}:\ t \approx 8336\ \mathrm{s} \end{cases} \tag{S24}$$

and,

$$\begin{cases} X_{gb} = 50\%:\ t \approx 253\ \mathrm{s} \\ X_{gb} = 75\%:\ t \approx 1263\ \mathrm{s} \\ X_{gb} = 90\%:\ t \approx 8336\ \mathrm{s} \end{cases} \tag{S25}$$

Text S5. Details of Materials and Ultrafast High-Temperature Sintering (UHS)

Plasma-atomized Cu powders (15-53 μm, 99.9% purity) were purchased from Bright Laser Technologies Co., Ltd. Zirconia powders ($Ce_{0.06}Zr_{0.94}O_2$ in mole ratio, average size ~2 μm, **fig. S1** and **Table S1**) were prepared by co-sintering method [30], and hereafter denoted as 6CZ. The Cu powders were co-milled with 20% 6CZ (in weight ratio) at 300 rpm for 12 h under high-purity argon using hardened steel vials and balls (QM-DY2 planetary mill). The as-milled composite powders (**fig. S2**) were then cold-pressed in a steel die (8 mm diameter) at 1.0 GPa for 1 min to form a green compact.

Ultrafast high-temperature sintering (UHS) was carried out in a custom-built setup inside a 99.99% Ar glove box (**Fig. 2**$a_1$). The green compact was embedded between carbon-fabric strips (PAN-based carbon tape, ~12 cm × 3 cm × 6.3 mm with a 2 cm slit, **Fig. 2**$a_2$) which serve as a Joule heating element. A DC power supply was used to drive a predefined current profile through the carbon fabric, resulting in resistive heating of the specimen (**Fig. 2**$a_3$). The current was ramped at ~1 A/s, producing a heating and cooling rate on the order of 100–200°C/s [31]. An infrared pyrometer (Raytek Raynger 3i Plus; measuring range: 700–3000°C; accuracy: ±10°C) was directed at the surface of the carbon heater to provide real-time temperature monitoring near the specimen. The instrument was focused on the upper surface of the carbon fabric in direct contact with the sample. Owing to the small thickness of the green bodies (<2 mm), the temperature difference between the top and bottom surfaces is expected to remain below 20°C after 5 s [32-34].

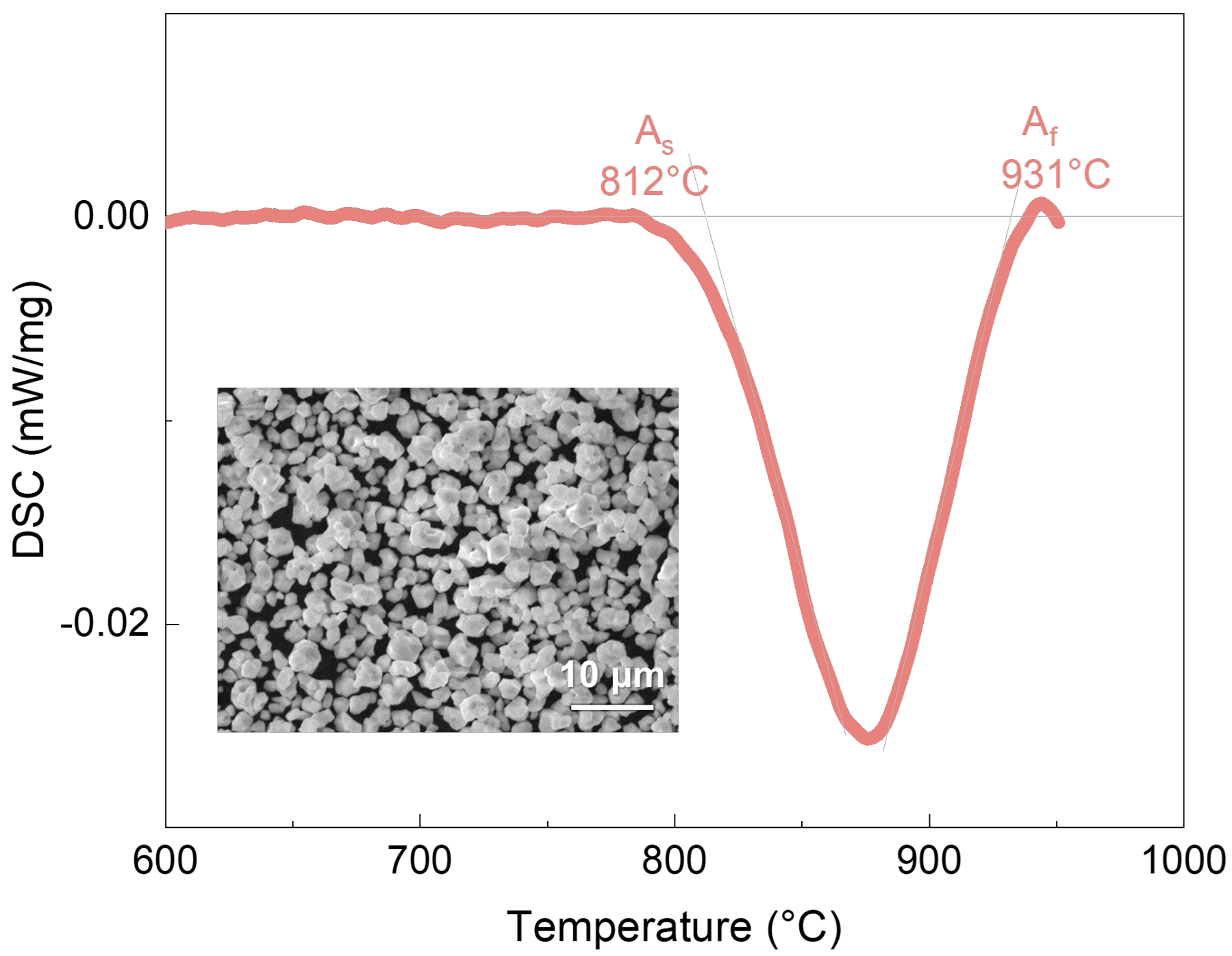

**Fig. S1**. Differential scanning calorimetry (DSC) curves of 6CZ particles, which was performed in argon with a heating/cooling rate of 10°C/min using $Al_2O_3$ crucibles. The inset shows the scanning electron microscopy (SEM) image of as-fabricated single crystalline 6CZ particles.

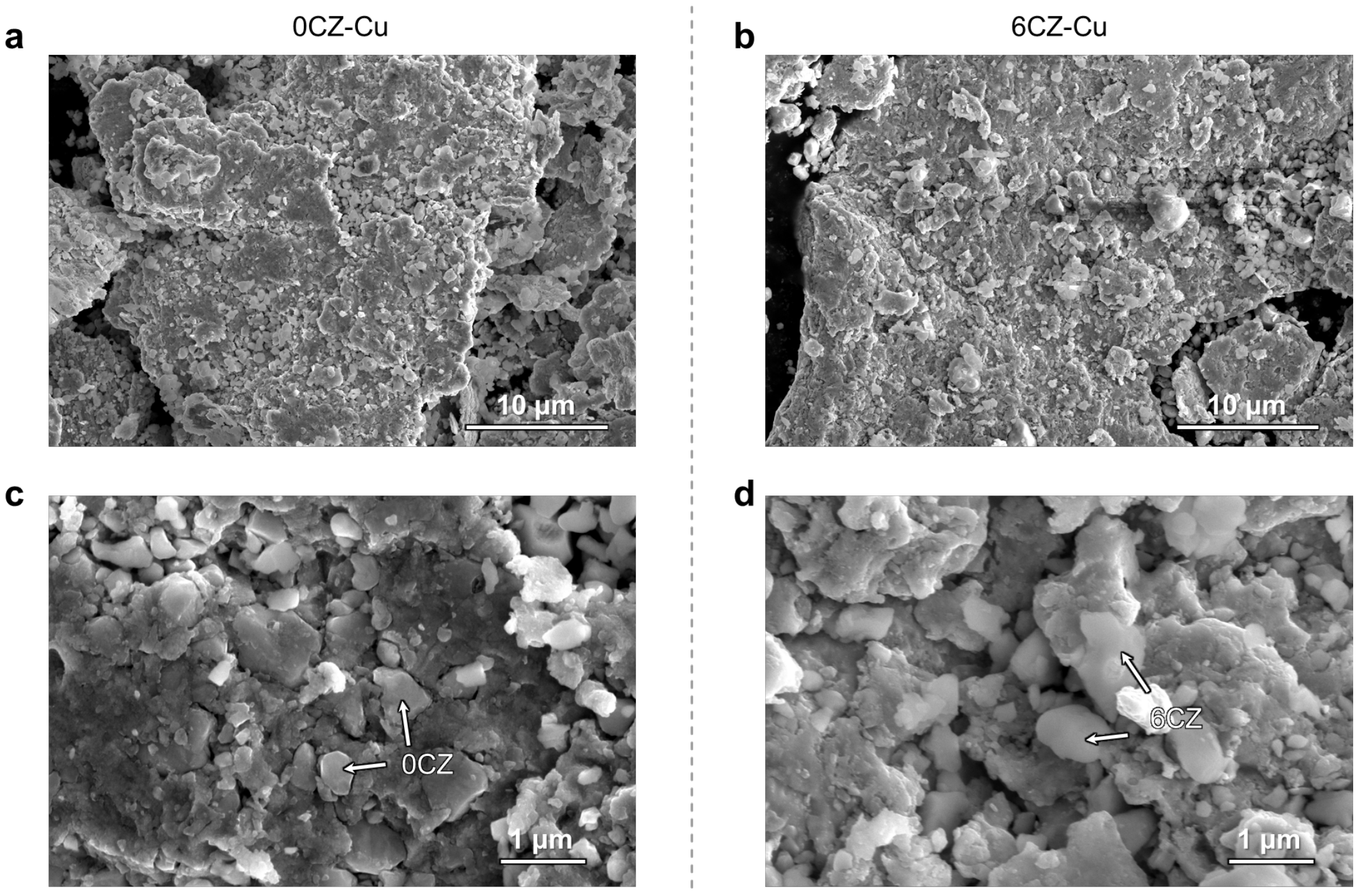

**Fig. S2**. SEM images of as-milled zirconia-copper composite powders. Low-magnified images of (a) 0CZ-Cu and (b) 6CZ-Cu; High-magnified images of (c) 0CZ-Cu and (d) 6CZ-Cu.

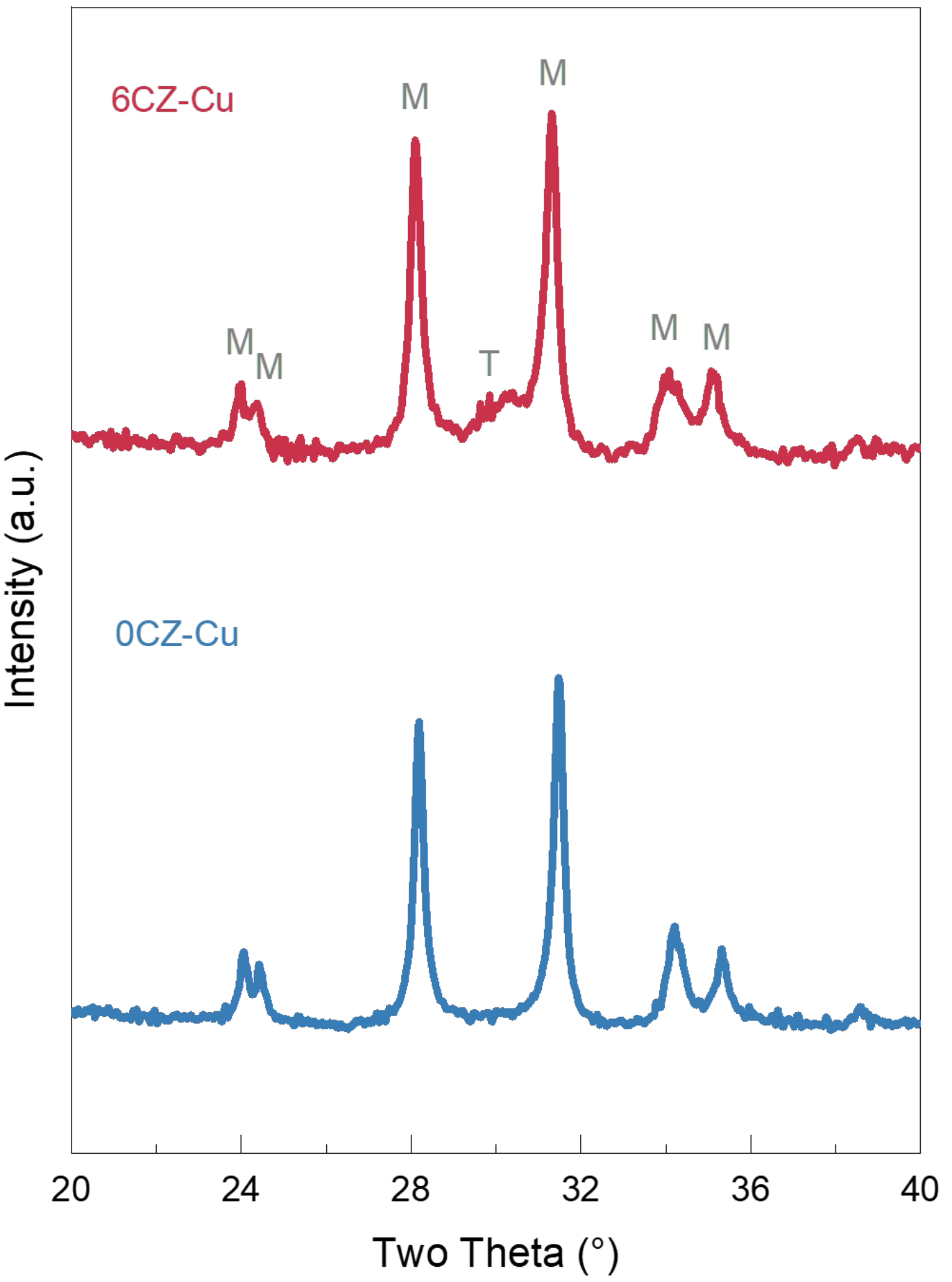

**Fig. S3**. X-ray diffraction (XRD) patterns of as-milled zirconia-copper composite powders.

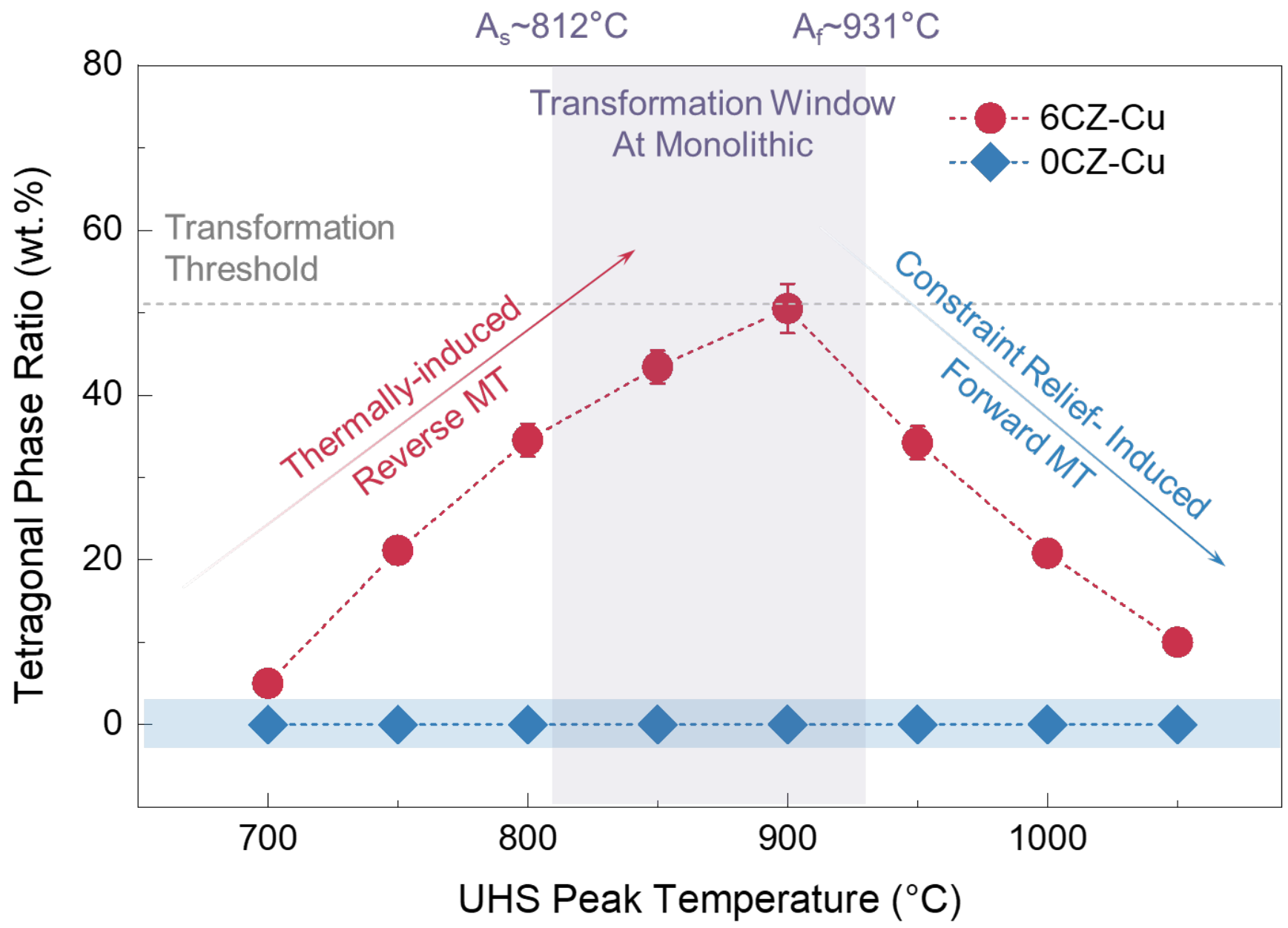

**Fig. S4**. XRD measured tetragonal phase weight ratio in 6CZ–Cu and 0CZ-Cu as a function of sintering peak temperature for UHS (20 s hold).

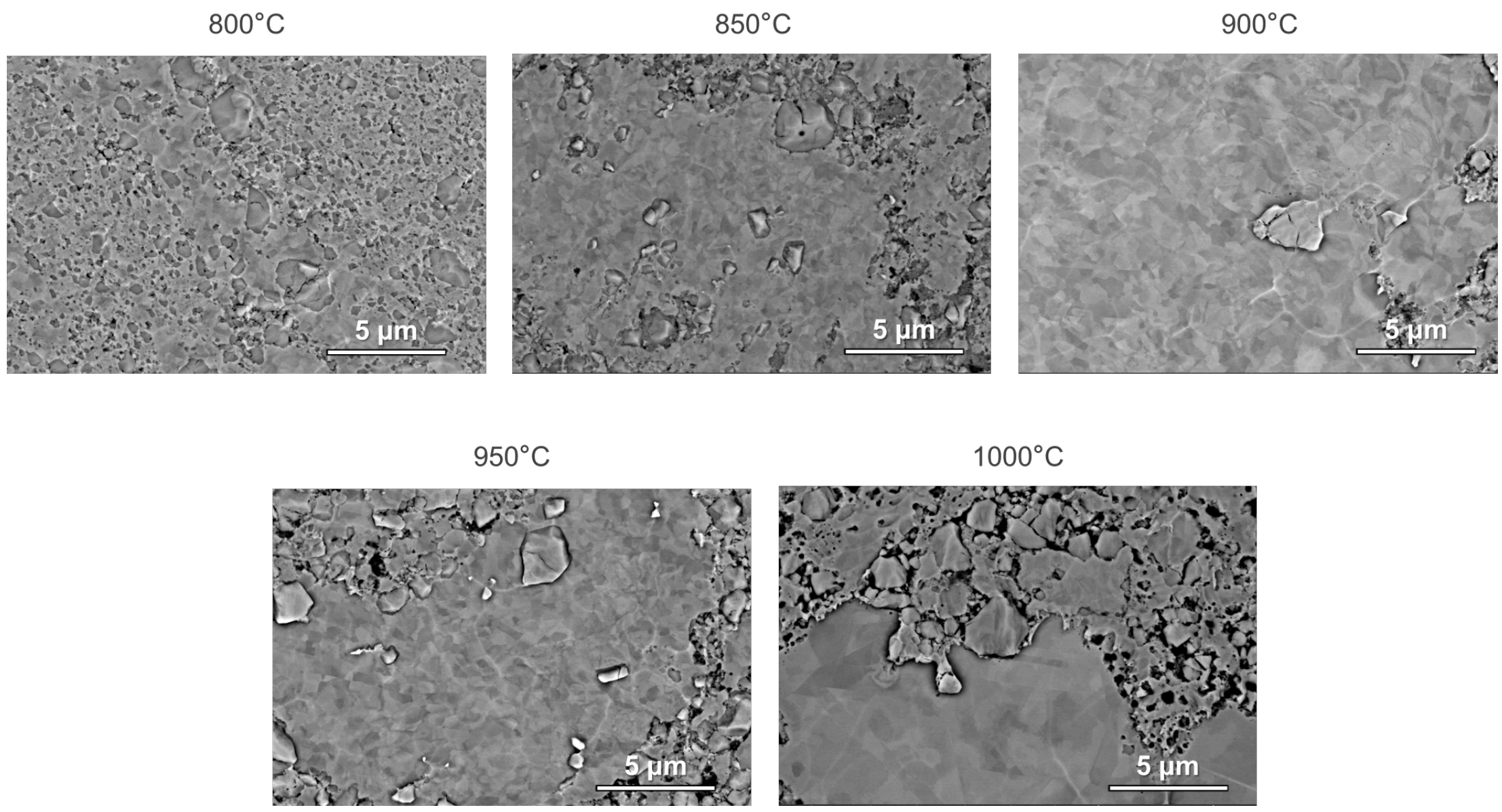

**Fig. S5**. Low-magnified SEM images of 6CZ–Cu cermets with varying UHS peak temperatures (20 s).

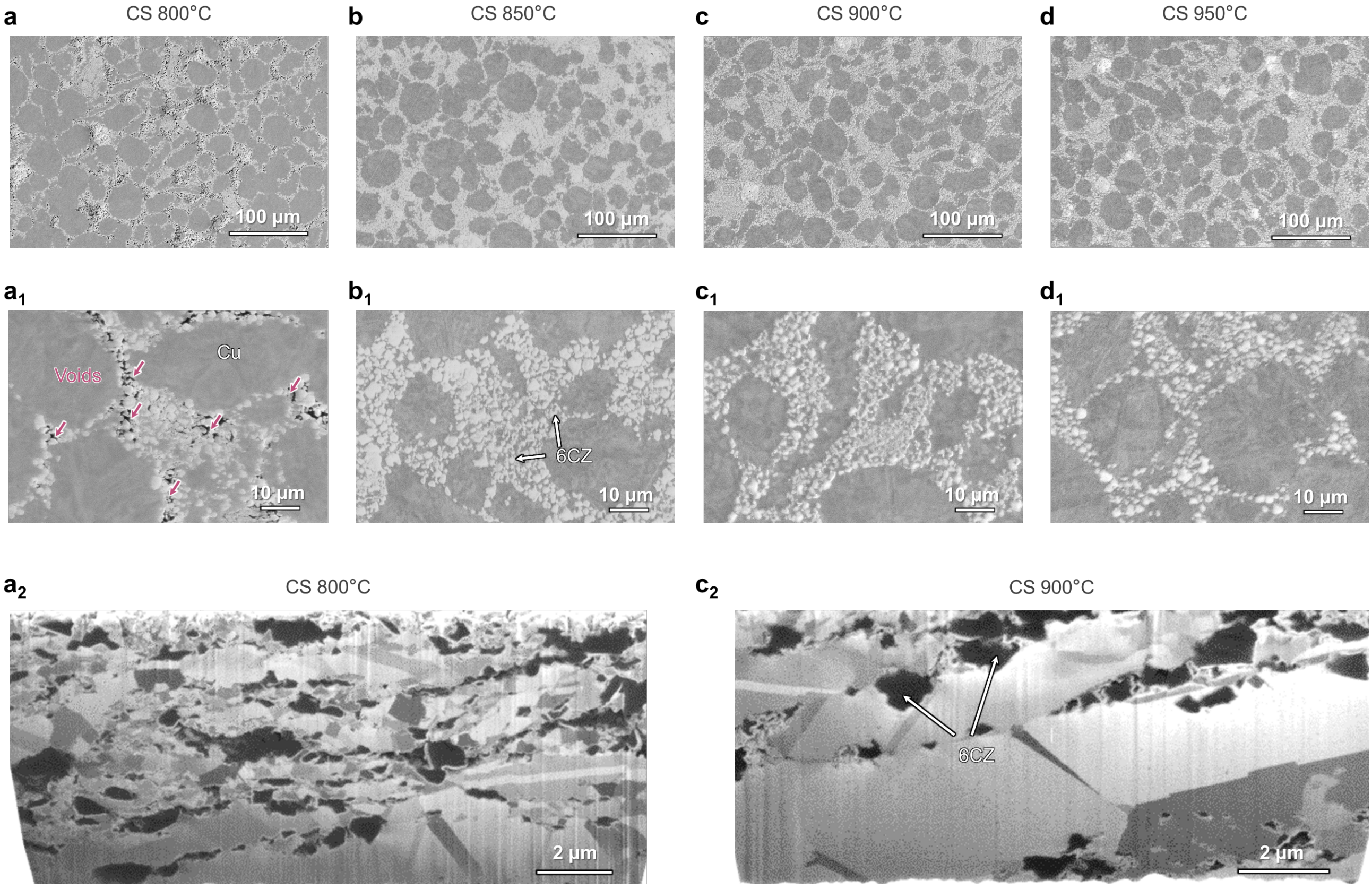

**Fig. S6**. Microstructures of 6CZ–Cu cermets with varying conventional sintering (CS) peak temperatures (1 h hold). (a–d) Low-magnified SEM images of cermets sintered at (a) 800°C, (b) 850°C, (c) 900°C, and (d) 950°C. The corresponding high-magnified SEM images are displayed in ($a_1$-$d_1$). The cross-sections of CS-treated 6CZ-Cu at ($a_2$) 800°C and ($c_2$) 900°C via focused ion beam (FIB).

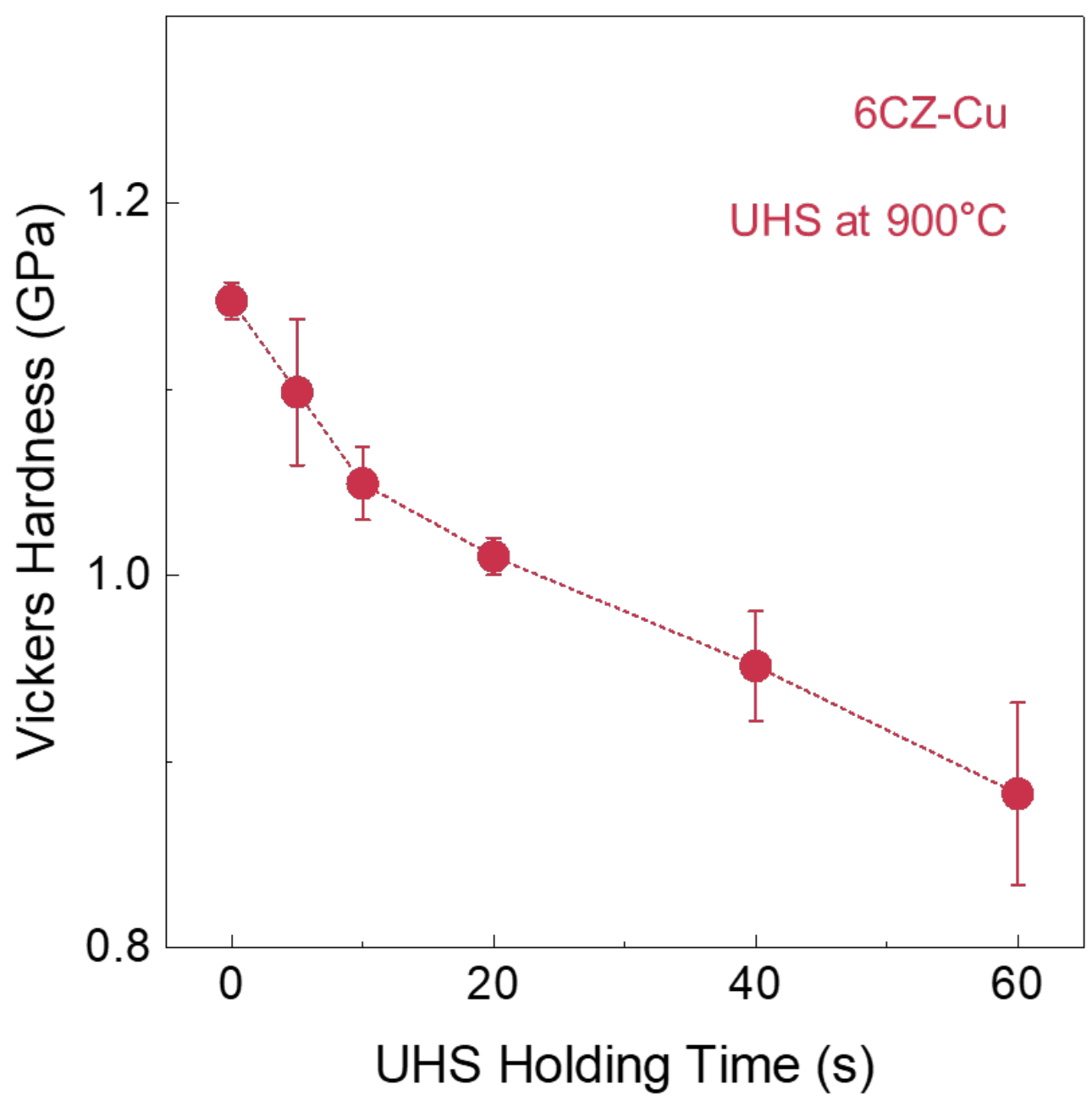

**Fig. S7**. Vickers Hardness for 6CZ–Cu as a function of UHS holding time (900°C).

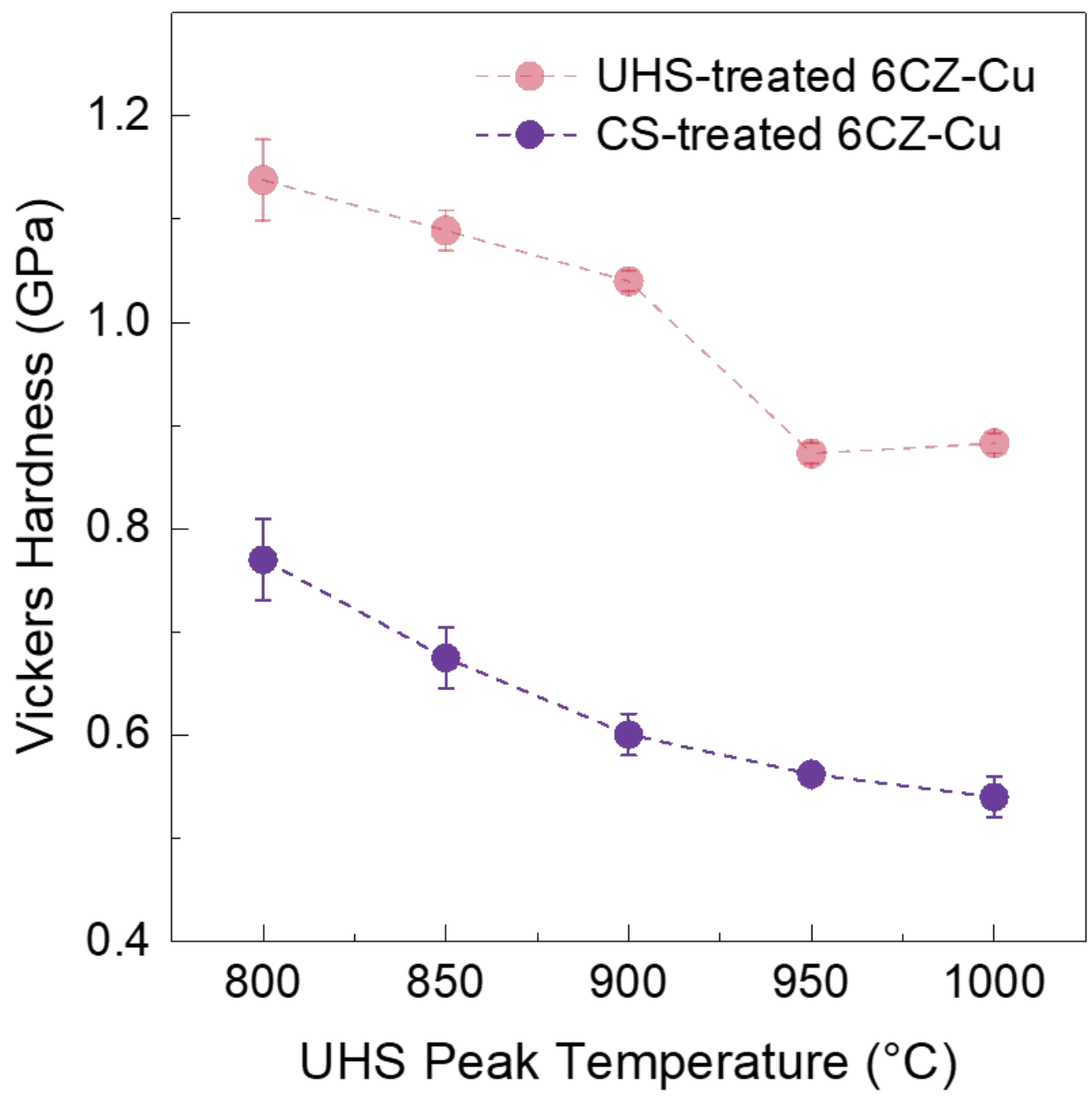

**Fig. S8**. Vickers Hardness for 6CZ–Cu as a function of CS peak temperatures (holding time 1 h). The hardness data from UHS-treated counterparts are shown for comparison.

**Table S1.**

Transformation temperature of zirconia-based phase-transforming ceramics*. (Average diameter of ~2 μm.)

| | $M_f$ (°C) | $M_s$ (°C) | $A_s$ (°C) | $A_f$ (°C) | $T_0$ (°C) |
|---|---|---|---|---|---|
| 0CZ | 991 | 1043 | 1150 | 1231 | 1104 |
| 6CZ | 582 | 728 | 812 | 931 | 763 |

Note: * $M_s$, $M_f$, $A_s$, $A_f$ refers to the martensite/austenite start/finish temperatures, respectively. $T_0$ is the average of $M_s$, $M_f$, $A_s$, and $A_f$.

**Table S2.**

Examples of austenite fraction transformed $X_{aus}$ and time with relative to temperature.

| | | **Ramping rate: 100 K/s** | **Ramping rate: 0.1 K/s** |
|---|---|---|---|
| **T/K** | $\boldsymbol{X_{aus}}$* | **t/s** | **t/s** |
| **873** | 0.0000 | 5.75 | 5750 |
| **1036 ($T_0$)** | 0.0000 | 7.38 | 7380 |
| **1073** | 0.3818 | 7.75 | 7750 |
| **1123** | 0.6773 | 8.25 | 8250 |
| **1173** | 0.8315 | 8.75 | 8750 |
| **1223** | 0.9121 | 9.25 | 9250 |

Note: * The transformed fraction under different matrix constraint may vary. Here, the calculation is based on the matrix constraint stress of -827 MPa obtained from our previous work in aluminum matrix [8].